\documentclass[twocolumn]{aastex63}

\usepackage{xspace}
\usepackage{float}

\newcommand{\pccm}{\ensuremath{\,\mathrm{cm}^{-3}}}	
\newcommand{\expo}[1]{\ensuremath{{\tau = #1\,\,\mathrm{mbarn}^{-1}}}}
\newcommand{\expogreater}[1]{\ensuremath{\tau \gtrsim #1\,\,\mathrm{mbarn}^{-1}}}
\newcommand{\n}[1]{\ensuremath{n = 10^{#1}\,\mathrm{cm}^{-3}}}
\newcommand{\metal}[1]{\ensuremath{{\left[ \mathrm{Fe} / \mathrm{H} \right] = #1 }}}
\newcommand{\FeH}{\ensuremath{\left[ \mathrm{Fe} / \mathrm{H} \right] }}
\newcommand{\el}[2]{\ensuremath{^{#1}\mathrm{#2}}}
\newcommand{\LP}{\textsc{LP625-44}\xspace}
\newcommand{\CS}{\textsc{CS31062-050}\xspace}
\newcommand{\CShightau}{\textsc{CS29497-030}\xspace}
\newcommand{\HElowtau}{\textsc{HE0336+0113}\xspace}

\newcommand{\Jstar}{\textsc{J052043}\xspace}
\newcommand{\pone}{Paper I\xspace}
\newcommand{\Msun}{\ensuremath{\,\mathrm{M}_{\odot}}}
\newcommand{\redchi}{\ensuremath{\chi^2_{\mathrm{red \,}}}}
\newcommand{\ndens}[2]{\ensuremath{ n  #2 10^{#1}\,\mathrm{cm}^{-3}  }}
\newcommand{\sqbr}[2]{ \ensuremath{\left[ \mathrm{#1} / \mathrm{#2} \right]} }
\newcommand{\Z}[1]{($Z=#1$)}

\graphicspath{{./}{figures/}}

\received{June 26, 2019}
\revised{October 7, 2019}
\accepted{October 17, 2019 for publication in ApJ}

\submitjournal{ApJ}

\shorttitle{The i process \& Pb abundances}
\shortauthors{M. Hampel et al.}

\begin{document}

\title{Learning about the intermediate neutron-capture process from lead abundances\footnote{This paper is dedicated to the celebration of the 100th birthday of Margaret Burbidge, in recognition of her outstanding contribution to nuclear astrophysics.}}

\correspondingauthor{Melanie Hampel}
\email{Melanie.Hampel@monash.edu}

\author[0000-0002-6963-5763]{Melanie Hampel}
\affiliation{Monash Centre for Astrophysics, School of Physics and Astronomy, Monash University, VIC 3800, Australia}

\author{Amanda I. Karakas}
\affiliation{Monash Centre for Astrophysics, School of Physics and Astronomy, Monash University, VIC 3800, Australia}

\author{Richard J. Stancliffe}
\affiliation{E. A. Milne Centre for Astrophysics, Department of Physics \& Mathematics, University of Hull, HU6 7RX, UK}

\author{Bradley S. Meyer}
\affiliation{Department of Physics and Astronomy, Clemson University, Clemson, SC, 29634-0978, USA}

\author{Maria Lugaro}
\affiliation{Konkoly Observatory, Research Centre for Astronomy and Earth Sciences, Hungarian Academy of Sciences, H-1121 Budapest, Hungary}
\affiliation{Monash Centre for Astrophysics, School of Physics and Astronomy, Monash University, VIC 3800, Australia}

\begin{abstract}

Lead (Pb) is predominantly produced by the slow neutron-capture process (s process) in asymptotic giant branch (AGB) stars. In contrast to significantly enhanced Pb abundances predicted by low-mass, low-metallicity AGB-models,  observations of Magellanic post-AGB stars show incompatibly low Pb abundances. Observations of carbon-enhanced metal-poor (CEMP) stars whose s-process enrichments are accompanied by heavy elements traditionally associated with the rapid neutron-capture process (r process) have raised the need for a neutron-capture process operating at neutron densities intermediate to the s and r process: the so-called i process.
We study i-process nucleosynthesis with single-zone nuclear-network calculations. Our i-process models can explain the heavy-element abundance patterns measured in Magellanic post-AGB stars including their puzzlingly low Pb abundances. Furthermore, the heavy-element enhancements in the post-AGB and CEMP-i stars, particularly their Pb abundance, allow us to characterise the neutron densities and exposures of the i process that produced the observed abundance patterns. We find that the lower-metallicity CEMP-i stars ($\left[ \mathrm{Fe} / \mathrm{H} \right] \approx -2.5$) have heavy-element abundances best matched by models with higher neutron densities and exposures ($\tau > 2.0 \, \mathrm{mbarn}^{-1}$) compared to the higher-metallicity post-AGB stars ($\left[ \mathrm{Fe} / \mathrm{H} \right] \approx -1.3$, $\tau < 1.3 \, \mathrm{mbarn}^{-1}$). This offers new constraints and insights regarding the properties of i-process sites and demonstrates that the responsible process operates on time scales of the order of a few years or less.

\end{abstract}


\section{Introduction} \label{sec:intro}

The large majority of elements heavier than iron are formed by the slow (s) and rapid (r) neutron-capture processes. These processes 
differ in their characteristic neutron density which determines how likely it is that an unstable neutron-rich isotope captures an additional neutron before decaying \citep{Burbidge1957, Meyer1994, Sneden2008}. 
The resulting abundance patterns for these processes shift as their conditions change opening up different neutron capture paths.

The s process operates at low neutron densities of \ndens{11}{\lesssim} and its main site is within asymptotic giant branch (AGB) stars \citep{Kappeler2011}. In this late evolutionary stage of low- and intermediate-mass stars with initial masses $\lesssim 8 \Msun$, an AGB star experiences thermal pulses (TP) and sometimes dredge-up events, which mix the material from the intershell to the stellar surface including carbon and s-process elements. In the intershell, conditions may allow for the formation of a small (in mass) region that is rich in \el{13}{C} (also referred to as a \el{13}{C} ``pocket"). Subsequent release of free neutrons via $\el{13}{C} \left( \alpha, \mathrm{n} \right) \el{16}{O}$ and their captures on Fe seed nuclei produce typical s-process elements, such as Sr, Y, Zr, Ba, La, Ce, and Pb.
Strong mass loss on the AGB releases these elements into the interstellar medium and makes AGB stars important contributors to the chemical evolution of galaxies. For details and reviews about AGB stars and the s process we refer the reader to \citet{Busso1999, Herwig2005, Karakas2014} and references therein.

In contrast, the r process operates at high neutron densities of \ndens{20}{\gtrsim} \citep{Meyer1994}. Typical elements produced by the r process include Eu, Os, Au and Pt. Different sites have been proposed to host the extreme conditions that are required for the r process, such as magneto-hydrodynamically driven supernovae \citep{Ono2012, Winteler2012, Nishimura2015, Nishimura2017} and neutron star mergers \citep{Lattimer1974, Meyer1989, Thielemann2017, Kilpatrick2017, Cote2018a}.

Observational evidence for another neutron-capture process intermediate to the s and r process comes from observations of carbon-enhanced metal-poor \citep[CEMP, ][]{Beers2005} stars with enhancements of Ba and Eu. Generally, it is believed that CEMP stars with Ba enhancements are formed in a binary system and that their C and s-process enhancements are the results of mass-transfer from an AGB companion. However, the additional Eu enhancements of some CEMP stars cannot be explained with this formation scenario and pollution from separate s- and r-process sites have been considered unsuccessfully \citep{Jonsell2006, Lugaro2012, Bisterzo2012, Abate2016}. It now seems likely that the observed abundance patterns are the result of a neutron-capture process that can occur in one single stellar site and operates at neutron densities intermediate to the s and r process ($n \approx 10^{14} - 10^{15} \pccm$) \citep[e.g.][]{Dardelet2014, Hampel2016}: the intermediate neutron-capture process \citep[i process,][]{Cowan1977, Malaney1986}.

Another example of objects with a puzzling heavy-element enrichment are post-AGB stars \citep[e.g.][]{van-Winckel2003} in the Small and Large Magellanic Clouds (SMC and LMC). From the inferred absolute luminosities it is possible to constrain the initial masses of the progenitors to $1 - 1.5 \Msun$ at metallicities between $-1 \leq \FeH \leq -1.3$ \citep{De-Smedt2012, van-Aarle2013, De-Smedt2014, De-Smedt2016}. 
Stellar models in this mass and metallicity range can produce high yields of s-process elements and due to fewer iron seeds at lower metallicities, the chain of neutron captures reaches all the way to the end of the neutron capture path at Pb  \citep{Gallino1998, De-Smedt2012}.
Measurements of the Pb abundance in post-AGB stars in the LMC and SMC are only able to provide upper limits. However, even these upper limits are lower than the high abundances predicted from AGB models with the required low progenitor mass and metallicity \citep{De-Smedt2012, De-Smedt2014, van-Aarle2013, Lugaro2015, Trippella2016}.

In order to reconcile the predicted and observed Pb abundances, uncertainties and parameters in AGB models have been studied, in particular with respect to the \el{13}{C} pocket responsible for the production of free neutrons. These attempts were not able to resolve the discrepancies \citep[e.g.][]{Lugaro2015, Trippella2016}. \citet{Lugaro2015} demonstrated that the full elemental pattern is incompatible with the s process and proposed that these abundance patterns might be better explained by i-process nucleosynthesis resulting from proton-ingestion episodes in the AGB progenitor. However, they did not explore this idea in detail with an appropriate nuclear network.

\citet[][from here on \pone]{Hampel2016} studied the equilibrium heavy-element abundance patterns characteristic for different neutron densities up to \n{15}. This deliberately excluded investigating Pb, which always gets produced as a function of time and does not reach equilibrium with Fe and the other heavy elements.
Via comparisons to the abundance patterns in CEMP stars \pone identified the i process as the best explanation for stars that show both Ba and Eu enrichments and renamed them CEMP-i stars (formerly known in the literature as CEMP-s/r and variations thereof).
However, additional insight into the total amount of captured neutrons and the time scale of the process is also essential for further interpretation.

This study aims at extending \pone by systematically studying the parameters that are responsible for shaping i-process abundance patterns and the production of Pb. This includes investigating the dependence on both the neutron density and the integrated neutron exposure, the latter as a proxy for the time evolution constrained by observed Pb abundances.
By comparing our results to the observed heavy-element abundances, particularly those of Pb measured in CEMP-i and Magellanic post-AGB stars, we aim to ($i$) show if the puzzlingly low Pb abundances can be attributed to the i process and ($ii$) constrain the neutron densities and exposures characteristic for the i process in these two types of objects with different metallicities. 
This is an important step towards identifying the site at which the i process operated in our observed samples. Moreover, linking the abundance patterns from CEMP-i and Magellanic post-AGB stars to the characteristic properties of the underlying neutron captures, we can infer observational constraints on the metallicity dependence of the i process.

We describe our method in \S\ref{sec:method} and show the results of our nuclear-network calculations in \S\ref{sec:nuc-net-results}. In \S\ref{sec:stars} we present the data sample of CEMP-i and post-AGB stars
that provides us the observational i-process probes for this
study. We compare our simulations to the observations in \S\ref{sec:comparison} with separate evaluations of the CEMP-i and post-AGB star comparisons in \S\ref{sec:CEMP-results} and \S\ref{sec:pAGB-results}, respectively. A discussion of these results is provided in \S\ref{sec:Discussion} complemented by a final summary and concluding remarks in \S\ref{sec:summary}.

\section{Method} \label{sec:method} 

As in \pone, we study the production of heavy elements by the i process by feeding constant neutron densities into a one-dimensional, single-zone nuclear-reaction network using \textit{NucNet Tools} \citep{Meyer2012} and the JINA Reaclib database \citep{Cyburt2010}. We model heavy-element production in conditions typical of the intershell region of a low-mass, low-metallicity AGB star in the presence of constant neutron densities of different magnitudes.
Temperatures and densities in the range 
$1.0\times10^{8}$\,K $ \leq T \leq 2.2\times10^{8}$\,K and $800$\,g\,cm$^{-3} \leq \rho 
\leq 3200$\,g\,cm$^{-3}$ were investigated, corresponding to the AGB-intershell conditions used by \citet{Stancliffe2011}. Compared to charged-particle reactions, neutron-capture reactions are less temperature sensitive. Consequently, we find that different temperatures and densities do not result in significant changes in the heavy-element abundance patterns. Therefore, in the following, we present one representative case with $T=1.5\times10^{8}$\,K and $\rho = 1600$\,g\,cm$^{-3}$. We note that the temperature and density significantly affect the activation of the neutron-producing charged-particle reactions, which are not studied in this work.

We use two different initial compositions for our calculations of possible i-process enrichment in the CEMP-i and post-AGB stars to account for the different metallicities of the observed objects. Such different initial compositions are used for consistency with the metallicities of the CEMP-i and post-AGB stars. However, for nuclear network calculations with imposed constant neutron densities, a different initial composition does not result in significantly different heavy-element abundance patterns. For the CEMP-i stars we use abundances from the intershell of a 1$M_{\odot}$-AGB model with metallicity \metal{-2.3} at the second thermal pulse from \citet{Lugaro2012}. For the post-AGB stars we use abundances from the intershell of a 1.7$M_{\odot}$-AGB model with metallicity \metal{-1.4} from \citet{Karakas2014a}, also at the second thermal pulse. In particular, these authors assumed an initial scaled-solar composition with solar abundances from \citet{Asplund2009}.

We expose the intershell material to constant neutron densities of $n = 10^{11}$, $10^{12}$, $10^{13}$, $10^{14}$, and $10^{15}\pccm$ to study the detailed abundance patterns created by neutron-capture processes. Note that the original definition of the i process by \citet{Cowan1977} refers to a neutron density of \n{15} and the lower neutron densities in our range up to \n{13} are known to occur as peak neutron densities in s-process simulations \citep[e.g.][]{Raiteri1991b, Gallino1998, The2007, Pignatari2010, Fishlock2014}.
This level of neutron density is typically only reached for a short amount of time in s-process models \citep[of the order of days, see e.g.][]{Fishlock2014} and can be the result of the activation of the \el{22}{Ne} source. Despite being short these intense neutron bursts can impact the heavy-element abundance pattern by opening up specific s-process branching points. However, these peak-s-process neutron densities do not produce the distinct i-process patterns that form when the bulk of the neutron irradiation occurs at the higher neutron density, as we model here.
It is not clear that the s- and i-process neutron densities are separated by a fixed boundary. Instead, we expect an overlap of the associated neutron-density regimes, depending on the actual physical i-process sites and the resulting neutron-capture time scales. In order to study the transition between ``typical'' s- and i-process abundance patterns it is important to cover all these neutron densities in a systematic study. 

We use the time-integrated neutron exposure $\tau$ to quantify the total amount of captured neutrons and the time scale of the responsible process:
\begin{eqnarray}
 \tau = \int n \,  v_T \, dt  = \int n \, \sqrt{2 \, k_B \, T / m_n}\, dt  \, ,
\end{eqnarray}
with neutron density $n$, thermal neutron velocity $v_T$, Boltzmann constant $k_B$, temperature $T$, and neutron mass $m_n$. Note that $\tau \propto t$ for our case of a constant neutron density. It is the combination of neutron density and exposure that allows us to infer the detailed operation of the neutron producing reactions and will ultimately constrain the thermodynamic properties and structural details of the production site.

In \pone all calculations were run until complete equilibrium was reached between the heavy elements lighter than Pb,
which was ensured by an unnaturally large exposure of \expo{495}. This means that a steady flow between neighbouring species was reached when the reaction rates of production and destruction (neutron-capture reactions and $\beta^{-}$ decays) reach equilibrium. Constant abundance ratios between species are created and result in heavy-element abundance patterns that are characteristic at each neutron density, and independent from the actual total neutron exposure.
However, Pb is only produced by neutron captures but not significantly destroyed and therefore does not reach equilibrium with Fe or the other heavy elements. 

Here we expand our previous study from \pone by exploring different neutron exposures in order to investigate the evolution of the elements at the end of the neutron-capture chain and Pb in particular.
In practice, we extract the simulated abundance patterns of the produced unstable, neutron-rich isotopes from the nuclear-network calculations at each time step and evolve their further evolution without additional neutron exposure for $t=10$\,Myr. This allows the long-lived unstable isotopes to decay; a longer decay time than 10 Myr only has a negligible effect on the observable elemental abundances. For each tested neutron density, this results in a series of time- (and thereby exposure-) dependent heavy-element patterns.

\section{Results of the Nuclear Network Calculations} \label{sec:nuc-net-results}

When modelling the production of heavy elements by neutron-capture reactions, isotopes with magic neutron numbers are of particular interest.
Due to low neutron-capture cross sections relative to their neighbouring isotopes, nuclei with magic neutron numbers act as bottlenecks on the neutron-capture path and are produced in larger quantities than their neighbouring isotopes. For the s process this gives rise to the Solar s-process element abundance pattern with the light s-process (ls) peak with representative elements such as Sr, Y and Zr (atomic numbers $Z=38$, 39, and 40, respectively); the heavy s-process (hs) peak with representative elements such as Ba, La, and Ce (atomic numbers $Z=56$, 57, and 58, respectively); and the Pb \Z{82} peak \citep[e.g.,][]{Sneden2008, Karakas2014}. In \pone we discussed how the i process operates, the resulting heavy-element abundance patterns, and how they differ from the typical s process. In the following section we focus on the time-dependence of the i-process nucleosynthesis and the production of Pb in particular.

\begin{figure}[htp]
\centering
\includegraphics[ width=0.45\textwidth]{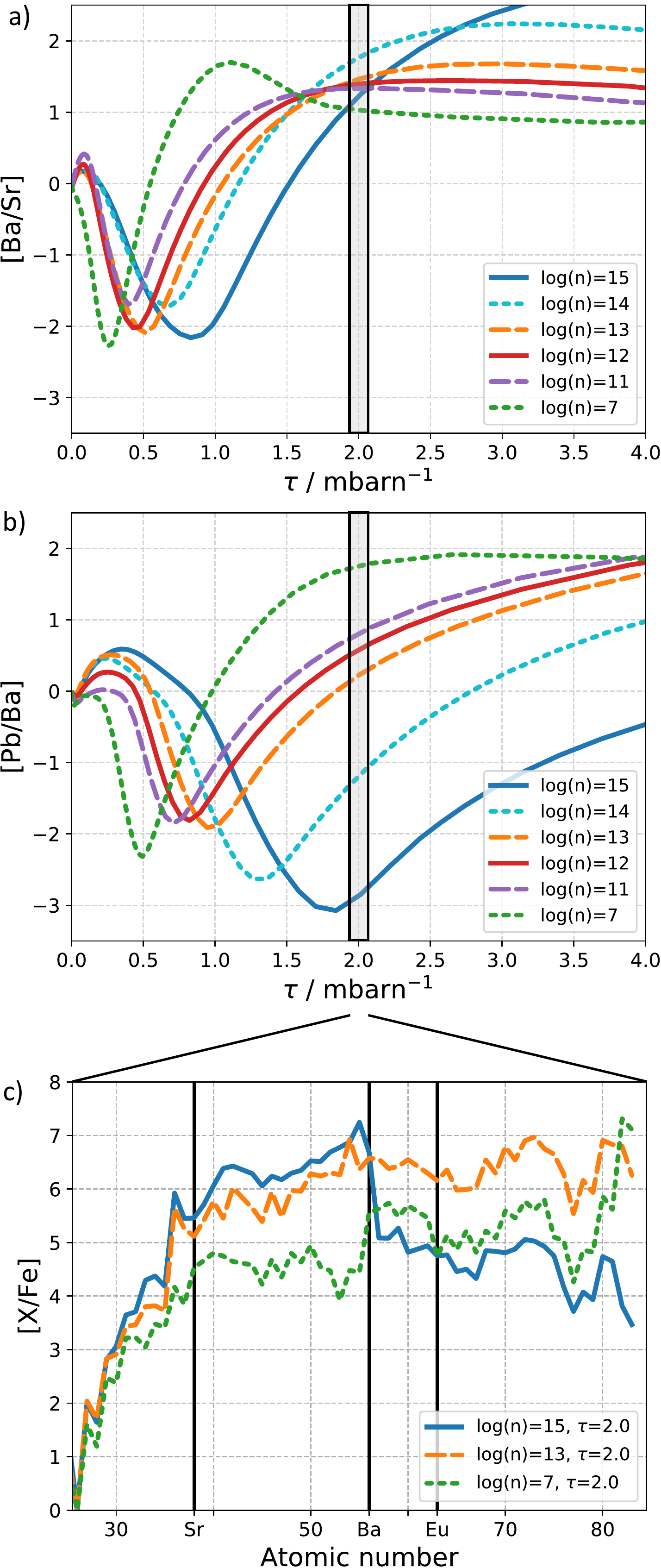}
\caption{Heavy element abundances of s- and i-process calculations with different neutron densities $n$:
a) and b) show the evolution of the relative abundances of the s-process peaks as a function of time integrated neutron exposure $\tau$. 
c) Shows the heavy-element abundance patterns in the simulations with different neutron densities $n$ at the same neutron exposure \expo{2.0}. The shown abundances refer to the final, post-decay abundances of the stable isotopes.
Note that these are predicted, undiluted abundances and cannot be directly compared to observed [X/Fe] ratios. Further dilution with solar-scaled material will decrease the [X/Fe] ratios to the levels observed in CEMP-i and post-AGB stars (see section \S\ref{sec:comparison}).}
\label{fig:Pb_Ba_Sr_production}
\label{fig:sqbr_Z_tau2.0}
\end{figure}

Figure \ref{fig:Pb_Ba_Sr_production} shows the evolution of the relative heights of the s-process peaks in our simulations, where we use Sr and Ba as representatives of the ls and hs peak, respectively. Since the material in our simulation is exposed to large neutron fluxes, the evolution of the s-process peaks is independent of the initial compositions. The shown abundances refer to the final, post-decay abundances of the stable nuclei. For comparison, a simulation with a typical s-process neutron density of \n{7} is also included in Figure \ref{fig:Pb_Ba_Sr_production}. To account for the different time scales on which neutron captures happen at different $n$, the Sr, Ba and Pb production ratios are shown as a function of the time-integrated neutron exposure $\tau$. All the simulations show an initial phase of decrease in [Ba/Sr] and [Pb/Ba] until they reach a minimum and begin to rise. 
For each simulation the rise of [Pb/Ba] starts after the rise in [Ba/Sr]. This shows that the main production of Sr, Ba and Pb follow one after the other in the respective order as expected. 

The minimum [Ba/Sr] and [Pb/Ba] for the different simulations shows that models with higher $n$ start their production phases of Ba and Pb at higher $\tau$. This shift shows a slow down of the heavy-element production in terms of integrated neutron exposure in comparison to the s process. 
As the i-process neutron-capture path moves away from the valley of stability more neutron-rich isotopes are produced. On average, these have lower neutron-capture cross sections than isotopes with fewer neutrons which are closer to the valley of stability. With a decrease of the average neutron-capture cross section an increase in neutron exposures is required for the production of heavier isotopes and Pb in particular.

In more detail, Figure \ref{fig:sqbr_Z_tau2.0} also shows the heavy-element abundance patterns of the simulations with $n = 10^{7}$,  $10^{13}$, and $10^{15}\pccm$ at the same integrated neutron exposure \expo{2.0}. From the different distributions of the produced heavy elements, it can be seen that the neutron-capture paths have progressed to very different stages, although the same amount of neutrons per neutron-capture cross section have been provided. 
The simulation with \n{15} shows a peak for the elements Xe, Cs, and Ba ($Z=54$, 55, and 56, respectively). The neutron-capture path just reaches the second bottleneck at the magic neutron number $N=82$ and it mainly produces magic isotopes of elements a few atomic numbers lighter than the traditional s-process elements. This results in a predominant production of the radioactive isotope \el{135}{I} (with decays to  \el{135}{Cs} and \el{135}{Ba}) instead of the stable magic s-process isotopes such as \el{138}{Ba}, \el{139}{La} and \el{140}{Ce}. Most of the heavy elements that have been 
produced at \n{15} due to this exposure of \expo{2.0} are lighter than Ba. Most notably, the main neutron-capture path has not reached the Pb peak yet and the stable decay products show $\sqbr{Pb}{Ba}=-2.9$.

In contrast, the neutron capture path of the simulation with \n{13} has passed the first bottlenecks at the magic neutron numbers $N=50$ and $N=82$ and has progressed to the third Pb peak. Pb is already produced at a level similar to that of the second magic neutron peak with $\sqbr{Pb}{Ba}=0.2$. 
The simulation with \n{7} shows the fastest progression in terms of neutron exposure, where the neutron-capture path has reached
the highest production of Pb with abundances as high as $\sqbr{Pb}{Fe}>7$ and $\sqbr{Pb}{Ba}>1.5$.

An important consequence is evident in Figure \ref{fig:Pb_Ba_Sr_production}:
the simulations at neutron densities with $n \geq 10^{11} \pccm$ cover regimes of [Ba/Sr]-[Pb/Ba] space that cannot be reached by standard s-process nucleosynthesis. With the exception of \n{15}, all the simulations show an almost identical ratio of $\left[ \mathrm{Ba} / \mathrm{Sr} \right]\approx 1.2$ at $\tau \approx 1.6 \,\mathrm{mbarn}^{-1}$. At the same neutron exposure, the simulations with different $n$ cover a range of more than 3 dex in the corresponding [Pb/Ba] abundance ratios. More generally, even at different exposures Figure \ref{fig:Pb_Ba_Sr_production} shows that the different neutron densities only result in relatively small changes in [Ba/Sr] compared to the larger ones in [Pb/Ba]. This gives rise to the possibility of explaining a range of heavy-element abundance patterns and in particular the different relative ratios of the ls, hs and Pb-peak using a range of neutron densities and exposures.

Figure \ref{fig:Pb_Ba_Sr_production} does not show the final phase of the abundance evolution, where the ratio between two elements becomes constant as a steady flow between them brings their abundances into equilibrium (studied in \pone). For \sqbr{Ba}{Sr} this happens at higher exposures than shown (\expogreater{10} for \n{7} and even higher for increasing $n$) and \sqbr{Pb}{Ba} does not reach an equilibrium at all. The second panel in Figure \ref{fig:Pb_Ba_Sr_production} suggests that such an equilibrium may be reached for \n{7} at \expogreater{2.5}, which is not the case. Instead, \sqbr{Pb}{Ba} enters a phase of decrease before increasing again at higher $\tau$. This is due to different production and destruction phases of Ba as it reaches equilibrium stages with other heavy elements, whereas the abundance of Pb keeps rising continuously and does not enter such steady-flow equilibria.

\section{Observational Sample} \label{sec:stars}

We consider observed heavy-element abundance patterns from CEMP-i and post-AGB stars. The sample of CEMP-i stars is adapted from the sample compiled by \citet{Abate2015b} and used in \pone. It comprises stars in the metallicity range of $-2.8 \leq \FeH \leq -1.8$ from the SAGA database \citep{Suda2008} and from \citet{Masseron2010} selected with the following criteria:
\begin{itemize}
\item $\sqbr{C}{Fe} > 1 \, ,$
\item $\sqbr{Eu}{Fe} > 1 $ and 
\item $\sqbr{Ba}{Eu} > 0 \, .$
\end{itemize}

We choose this CEMP-i definition for consistency with the definition of CEMP-rs from \citet{Masseron2010} and CEMP-s/r from \citet{Abate2015b}. Note that some authors prefer more restrictive definitions by including an extra constraint on the upper limit on the relative abundance ratios of s- to r-elements, e.g. $\sqbr{Ba}{Eu} < 0.5 $ \citep{Beers2005} or $\sqbr{La}{Eu} < 0.4 $ \citep{Bisterzo2012}. However, this restriction results in the misclassification of some stars as CEMP-s even though the heavy-element abundances are inconsistent with s-process nucleosynthesis and are better matched by i-process predictions (\pone). 

Out of the 20 stars from \citet{Abate2015b} we select the 16 stars with measured Pb abundances. We complement the sample with an additional 8 stars from the SAGA database \citep{Suda2008}, the JINABase \citep{Abohalima2018}, \citet{Masseron2010}, and \citet{Bisterzo2012}. These additional stars were selected because they fulfil the criteria above and have a measurement or upper limit of the Pb abundance. They extend the metallicity range of our sources down to $\FeH = -3$. 
For consistency with the compilation by \citet{Abate2015b} we adopt the same handling of multiple abundance and uncertainty measurements. 
We refer the reader to \citet{Abate2015b} for details regarding the data selection. Details of the 24 CEMP-i stars in our sample are listed in Table \ref{tab:data_sample}.

To investigate the puzzlingly low Pb abundances found in post-AGB stars, we consider the measured abundances and upper limits from five post-AGB stars in the LMC and SMC \citep{De-Smedt2012,van-Aarle2013, De-Smedt2014, De-Smedt2015}. 
We focus on the five post-AGB stars that show characteristics of being intrinsically enriched and exclude J053253, which instead has been suggested as a candidate for a young stellar object \citep{van-Aarle2013, Kamath2015}. Due to its high metallicity of \metal{-0.5} we also exclude J051213.
This makes our sample compatible with the study by \citet{Lugaro2015}. Additionally, we include the two most metal-poor Galactic post-AGB stars from \citet{De-Smedt2016} which have $\FeH < -0.7$. A summary of the 7 post-AGB stars selected for this study is given in Table \ref{tab:data_sample} where we show the full designations of the post-AGB stars considered in this study.

\citet{De-Smedt2016} concluded that their upper limits for the Pb abundances of stars with ${\FeH > -0.7}$ are compatible with predictions from AGB models. While the full abundance patterns were not tested for each individual star, it becomes evident from the \sqbr{Pb}{hs} and \sqbr{Pb}{ls} ratios (shown in their Figure 14) that \metal{-0.7} is the boundary below which the Pb discrepancy arises.

All together, this gives us a sample of 31 stars with Pb abundances or upper limits thereof, where we have 13 new i-process candidates and 16 stars that have been confirmed in \pone to exhibit an i-process signature. Our sample spans a total metallicity range of ${-0.8 \geq \FeH \geq -3.0}$.

\begin{table*}[!ht]
  \centering
  \caption{The data sample. Temperatures, surface gravities and selected chemical properties of the studied 24 CEMP-i and 7 post-AGB stars.} 
    \begin{tabular}{cccccccccc}
\toprule 
ID & $T_{\text{eff}}$  (K) & $\log g$ & [Fe/H] & [C/Fe] & [Ba/Fe] & [La/Fe] & [Eu/Fe] & [Pb/Fe] & Reference \\ 
\tableline 
\vspace{-5pt} & & & & & & & & & \\
\textbf{CEMP-i stars} & & & & & & & & & \\
BS16080-175 & 6240 & 3.7 & $-1.9\pm0.2$ & $1.8\pm0.2$ & $1.5\pm0.1$ & $1.6\pm0.1$ & $1.1\pm0.1$ & $2.7\pm0.1$ & 1 \\ 
BS17436-058 & 5690 & 2.7 & $-1.8\pm0.1$ & $1.6\pm0.2$ & $1.7\pm0.1$ & $1.6\pm0.2$ & $1.2\pm0.1$ & $2.3\pm0.1$ & 1 \\ 
CS22183-015 & 5540 & 3.2 & $-2.9\pm0.3$ & $2.3\pm0.2$ & $2.0\pm0.2$ & $1.7\pm0.2$ & $1.6\pm0.2$ & $2.9\pm0.1$ & 1, 14, 15, 30 \\ 
CS22887-048 & 6500 & 3.5 & $-1.8\pm0.2$ & $1.7\pm0.3$ & $1.9\pm0.2$ & $1.8\pm0.1$ & $1.5\pm0.2$ & $3.5\pm0.1$ & 1, 30 \\ 
CS22898-027 & 6110 & 3.7 & $-2.3\pm0.2$ & $2.0\pm0.4$ & $2.3\pm0.3$ & $2.3\pm0.2$ & $2.0\pm0.2$ & $2.9\pm0.2$ & 4, 5, 28, 32 \\ 
CS22948-027 & 4800 & 1.8 & $-2.5\pm0.3$ & $2.4\pm0.4$ & $2.4\pm0.4$ & $2.3\pm0.2$ & $1.9\pm0.2$ & $2.7\pm0.1$ & 7, 9 \\ 
CS29497-030 & 6966 & 4.0 & $-2.6\pm0.3$ & $2.5\pm0.2$ & $2.4\pm0.5$ & $2.1\pm0.2$ & $1.8\pm0.5$ & $3.6\pm0.5$ & 25, 36, 37, 35 \\ 
CS29497-034 & 4850 & 1.6 & $-3.0\pm0.3$ & $2.7\pm0.2$ & $2.2\pm0.1$ & $2.2\pm0.1$ & $1.9\pm0.1$ & $2.9\pm0.1$ & 7, 9 \\ 
CS29526-110 & 6500 & 3.2 & $-2.4\pm0.2$ & $2.3\pm0.4$ & $2.1\pm0.2$ & $1.8\pm0.2$ & $1.8\pm0.2$ & $3.4\pm0.2$ & 4, 5 \\ 
CS31062-012 & 6099 & 4.2 & $-2.8\pm0.8$ & $2.3\pm0.4$ & $2.1\pm0.2$ & $2.0\pm0.3$ & $1.6\pm0.3$ & $2.5\pm0.2$ & 3, 4, 5, 8, 33 \\ 
CS31062-050 & 5500 & 2.7 & $-2.4\pm0.2$ & $1.8\pm0.3$ & $2.5\pm0.2$ & $2.1\pm0.2$ & $1.8\pm0.2$ & $2.8\pm0.1$ & 5, 6, 26, 28 \\ 
HD187861 & 4960 & 2.0 & $-2.4\pm0.5$ & $2.0\pm0.2$ & $1.9\pm0.3$ & $2.1\pm0.3$ & $1.3\pm0.2$ & $3.1\pm0.3$ & 29, 39 \\ 
HD209621 & 4450 & 1.5 & $-2.0\pm0.3$ & $1.3\pm0.3$ & $1.8\pm0.3$ & $2.2\pm0.3$ & $1.6\pm0.3$ & $1.9\pm0.3$ & 22, 31 \\ 
HD224959 & 5050 & 1.9 & $-2.1\pm0.2$ & $1.8\pm0.2$ & $2.2\pm0.2$ & $2.2\pm0.2$ & $1.7\pm0.1$ & $3.1\pm0.2$ & 29, 39 \\ 
HE0143-0441 & 6305 & 4.0 & $-2.4\pm0.2$ & $2.0\pm0.2$ & $2.4\pm0.2$ & $2.0\pm0.2$ & $1.7\pm0.2$ & $3.4\pm0.3$ & 13, 14 \\ 
HE0243-3044 & 5400 & 3.2 & $-2.6\pm0.2$ & $2.4\pm0.3$ & $2.0\pm0.2$ & $2.5\pm0.2$ & $2.0\pm0.2$ & $3.1\pm0.2$ & 23 \\ 
HE0336+0113 & 5700 & 3.5 & $-2.8\pm0.3$ & $2.5\pm0.1$ & $2.6\pm0.3$ & $2.0\pm0.2$ & $1.3\pm0.2$ & $<2.3$ & 14, 15 \\ 
HE0338-3945 & 6161 & 4.1 & $-2.5\pm0.2$ & $2.1\pm0.2$ & $2.4\pm0.2$ & $2.3\pm0.2$ & $2.0\pm0.2$ & $3.0\pm0.1$ & 10, 27 \\ 
HE0414-0343 & 4863 & 1.2 & $-2.3\pm0.1$ & $1.4\pm0.3$ & $1.9\pm0.1$ & $1.6\pm0.1$ & $1.3\pm0.3$ & $2.3\pm0.1$ & 24 \\ 
HE1305+0007 & 4655 & 1.5 & $-2.2\pm0.3$ & $2.1\pm0.4$ & $2.6\pm0.5$ & $2.8\pm0.3$ & $2.2\pm0.3$ & $2.6\pm0.3$ & 11, 21 \\ 
HE1405-0822 & 5220 & 1.7 & $-2.4\pm0.1$ & $1.9\pm0.1$ & $2.0\pm0.2$ & $1.5\pm0.2$ & $1.6\pm0.2$ & $2.3\pm0.2$ & 16 \\ 
HE2148-1247 & 6380 & 3.9 & $-2.4\pm0.2$ & $2.0\pm0.2$ & $2.4\pm0.1$ & $2.3\pm0.2$ & $2.0\pm0.1$ & $3.1\pm0.2$ & 12 \\ 
HE2258-6358 & 4900 & 1.6 & $-2.7\pm0.1$ & $2.4\pm0.1$ & $2.3\pm0.1$ & $1.9\pm0.1$ & $1.7\pm0.1$ & $2.6\pm0.5$ & 34 \\ 
LP625-44 & 5500 & 2.6 & $-2.8\pm0.3$ & $2.3\pm0.2$ & $2.8\pm0.2$ & $2.6\pm0.3$ & $1.9\pm0.3$ & $2.6\pm0.2$ & 2, 3, 5, 6, 33 \\ 
\vspace{-5pt} & & & & & & & & & \\
\textbf{post-AGB stars} & & & & & & & & & \\
IRAS07134+1005 & 7250 & 0.5 & $-0.9\pm0.2$ & $1.1\pm0.2$ & $1.8\pm0.3$ & $1.8\pm0.2$ & $0.6\pm0.3$ & $<1.7$ & 20 \\ 
IRAS22272+5435 & 5750 & 0.5 & $-0.8\pm0.1$ & $1.0\pm0.1$ & ... & $2.2\pm0.1$ & $1.1\pm0.1$ & $<1.4$ & 20 \\ 
J004441.04-732136.4 & 6250 & 0.5 & $-1.3\pm0.3$ & $1.6\pm0.4$ & ... & $2.7\pm0.3$ & $1.9\pm0.2$ & $<2.3$ & 17, 18 \\ 
J050632.10-714229.8 & 6750 & 0.0 & $-1.2\pm0.2$ & $1.1\pm0.1$ & $1.2\pm0.3$ & $1.4\pm0.2$ & $0.5\pm0.3$ & $<1.2$ & 18, 38 \\ 
J051848.86-700246.9 & 6000 & 0.5 & $-1.0\pm0.1$ & $1.1\pm0.2$ & ... & $2.4\pm0.2$ & $1.3\pm0.2$ & $<1.6$ & 19 \\ 
J052043.86-692341.0 & 5750 & 0.0 & $-1.1\pm0.2$ & $1.2\pm0.2$ & ... & $1.9\pm0.2$ & $1.2\pm0.2$ & $<1.1$ & 18, 38 \\ 
J053250.69-713925.8 & 5500 & 0.0 & $-1.2\pm0.1$ & $1.5\pm0.2$ & ... & $1.9\pm0.3$ & $1.1\pm0.2$ & $<1.4$ & 18, 38 \\ 
\tableline 
\end{tabular}%

\tablerefs{(1) \citealt{Allen2012}; (2) \citealt{Aoki2000}; (3) \citealt{Aoki2001}; (4) \citealt{Aoki2002a}; (5) \citealt{Aoki2002b}; (6) \citealt{Aoki2006}; (7) \citealt{Aoki2007}; (8) \citealt{Aoki2008}; (9) \citealt{Barbuy2005}; (10) \citealt{Barklem2005}; (11) \citealt{Beers2007}; (12) \citealt{Cohen2003}; (13) \citealt{Cohen2004}; (14) \citealt{Cohen2006}; (15) \citealt{Cohen2013}; (16) \citealt{Cui2013}; (17) \citealt{De-Smedt2012}; (18) \citealt{De-Smedt2014}; (19) \citealt{De-Smedt2015}; (20) \citealt{De-Smedt2016}; (21) \citealt{Goswami2006}; (22) \citealt{Goswami2010}; (23) \citealt{Hansen2015}; (24) \citealt{Hollek2015}; (25) \citealt{Ivans2005}; (26) \citealt{Johnson2004}; (27) \citealt{Jonsell2006}; (28) \citealt{Lai2007}; (29) \citealt{Masseron2010}; (30) \citealt{Masseron2012}; (31) \citealt{Matrozis2012}; (32) \citealt{McWilliam1995}; (33) \citealt{Norris1997}; (34) \citealt{Placco2013}; (35) \citealt{Roederer2014}; (36) \citealt{Sivarani2004}; (37) \citealt{Sneden2003b}; (38) \citealt{van-Aarle2013}; (39) \citealt{van-Eck2003b}}
  \label{tab:data_sample}%
  \vspace{20cm}
\end{table*}%

\section{Comparison to observations} \label{sec:comparison}

To compare calculated heavy-element abundances to measured surface abundances of stars we assume some dilution with material of scaled-solar composition. Considering that we expect the i-process to operate only in a small region of the star, it is justified to expect that further evolution, e.g., dredge-up episodes in the AGB phase or mass-transfers onto a companion star, will dilute the i-process material. For this we assume a solar-scaled heavy-element distribution of the dilution material and compute the final abundances $X$:
\begin{eqnarray}
X = X_i \times \left( 1-d\right) + X_{\odot} \times d\, , 
\end{eqnarray} 
where $X_i$ is the calculated i-process abundance of each element, $X_{\odot}$ the solar-scaled abundance\footnote{For this work and the presented results we use the solar abundances from \citet{Asplund2009}.} and $d$ a dilution factor, which is a free parameter\footnote{Physically, we motivate this dilution by combining unprocessed material with the i-processed material. However, we do not consider particular physical processes in detail, such as specific mixing mechanisms or mass-transfer scenarios. Therefore we refrain from a quantitative interpretation of the dilution parameter.} with a value between 0 and 1. 
We compare the modelled abundances at different $n$ to the sample of observed surface abundances of CEMP-i and post-AGB stars by varying the free parameters $\tau$ and $d$ in order to find the minimal $\chi^2$:
\begin{eqnarray}
\chi^2 = \sum_{Z} \frac{\left( \left[ X_{Z}/\text{Fe}\right]_{\text{obs}} - \left[ X_{Z}/\text{Fe}\right]_{\text{mod}}\right) ^2}{\sigma_{Z,\text{obs}}^2} \, ,
\end{eqnarray}
where $\left[ X_{Z}/\text{Fe}\right]_{\text{obs}}$ and $\left[ X_{Z}/\text{Fe}\right]_{\text{mod}}$ are the observed and modelled abundances, respectively, of the element with atomic number $Z$ and $\sigma_{Z,\text{obs}}$ is the observational error of $\left[ X_{Z}/\text{Fe}\right]_{\text{obs}}$. For these calculations, only the abundances of the heavy elements with $ Z > 30 $ are taken into account. Furthermore, for the post-AGB stars in the LMC and SMC, the observed Pb abundances are only upper limits and consequently cannot be considered in the $\chi^2$ analysis of the fits. To put a special emphasis on the challenge that these Pb observations pose to nucleosynthesis models, we only consider models that do not produce Pb abundances above the observed upper limits.

In our analysis we use the absolute $\chi^2$ to identify the best fit for each star. 
A standard approach to compare the goodness of fits is using the reduced $\redchi = \chi^2 / \nu$, where $\nu$ is the number of degrees of freedom for the fit, which can only be reliably determined in the case of linear models without priors. Moreover, the observational data do not have uncorrelated Gaussian errors, which is why the statistical interpretation of \redchi does not apply here. The given $\sigma_{Z,\text{obs}}$ still give useful measures of how to weight different elemental abundances for a fit to an individual star; however, it is not meaningful to derive the goodness of the fits and uncertainties of the determined fit parameters based on $\chi^2$ statistics.

Figure \ref{fig:Pb_Ba_Sr_production} indicates that the fit parameters $n$ and $\tau$ are correlated, where models with higher $n$ also systematically require higher $\tau$ to produce the same abundance level. Moreover, Figure \ref{fig:Pb_Ba_Sr_production} suggests a certain degeneracy of the parameters where a different set of $n$ and $\tau$ can result in similar abundance ratios. In the absence of a meaningful \redchi measure to analyse the uncertainties and correlations of our parameters, we concentrate on presenting the best fitting models for each star. In some cases, almost equally good fits can be obtained from two models with different neutron densities, which we identify based on a 10\%-variation of the minimal $\chi^2$. We then also show the alternative fit to give an impression of how constrained the parameters are.

\subsection{Comparison to carbon-enhanced metal-poor stars} \label{sec:CEMP-results}

\begin{figure}[t!bp]
\centering
\includegraphics[ width=0.49\textwidth]{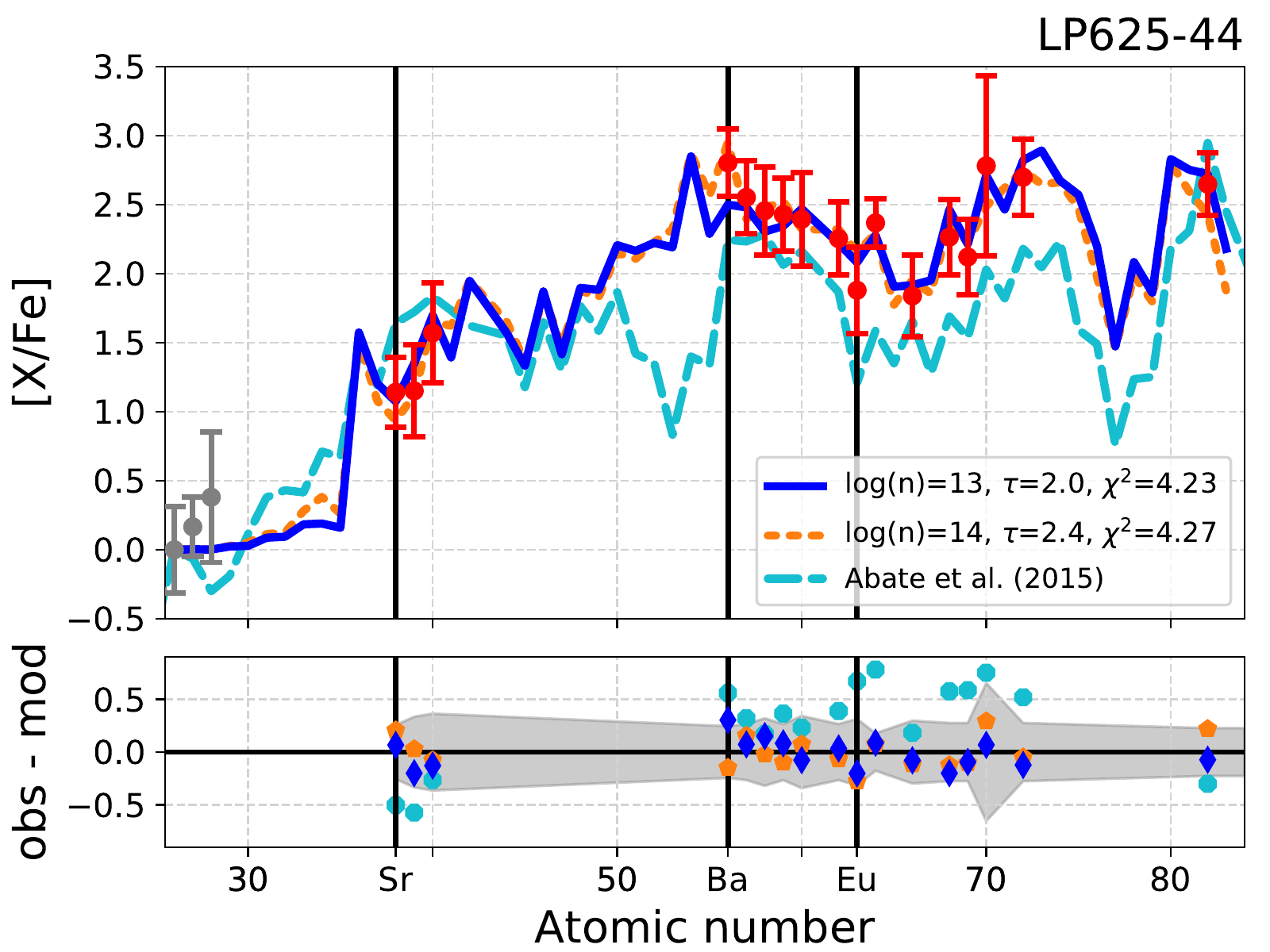} 
\caption{Best fits of the heavy-element abundance pattern of CEMP-i star \LP. The blue and orange line show the best fitting models with \n{13} and \n{14}, respectively, which describe the observed abundances almost identically well based on the $\chi^2$ value. For comparison the best fit of \citet{Abate2015b} is shown in cyan. The best fitting s-process models with initial r-process enhancement can be found in Fig. 31 of \citet{Bisterzo2012}.}
\label{fig:LPstar}
\end{figure}

We begin comparing our simulated abundance patterns to observations of the CEMP-i star \LP. This star has a metallicity of \metal{-2.75} and was discussed extensively in \pone. Here we refit the elemental composition of \LP using the method outlined in \S\ref{sec:method} which now includes the Pb abundance and an exposure dependence in the fit. Figure \ref{fig:LPstar} shows the two best fits of the modified method, which come from the simulations with \n{13} and \n{14}. For comparison, the best fit from \citet{Abate2015b} is also shown, which represents standard AGB evolution with s-process nucleosynthesis and binary interactions. We find the neutron density which reproduces the observed abundances of \LP best is \n{13}, which is lower than the previous result of \n{14} from \pone. However, based on the corresponding values of $\chi^2=4.23$ and $\chi^2=4.27$, respectively, these fits are almost equally as good to describe the observed abundances. Moreover, as in \pone a neutron density of \n{15} results in a significantly worse fit with $\chi^2=13.1$.

Our modified method shows that the neutron captures must occur over a relatively short time scale in order to match the data without overproducing Pb: for \LP, exposures as low as \expo{2.0} and \expo{2.4} at \n{13} and \n{14}, respectively, result in the best fits to the observations and correspond to time scales of 15 and 1.7 days, respectively.

In contrast to the good fits from the i-process simulations, Figure \ref{fig:LPstar} also shows that s-process simulations (\citealp[data from][]{Abate2015b}; \citealp[see also Figure 31 from][]{Bisterzo2012}) overproduce the Pb abundance in particular compared to the hs elements which are simultaneously underproduced. A less efficient s process, which is an assumption needed to better match the Pb observation, would also reduce the predicted hs abundances and hence even further increase the discrepancy that the hs measurements pose. 
It was already concluded in \pone that the observations of \LP show an abundance pattern that is best explained by i-process nucleosynthesis. The present additional consideration of the Pb enhancement of \LP, especially in comparison to the high enhancements of the elements with atomic numbers between $56 \leq Z \leq 72$, further strengthens this conclusion.

\begin{figure}[t!bp]
\centering
\includegraphics[ width=0.49\textwidth]{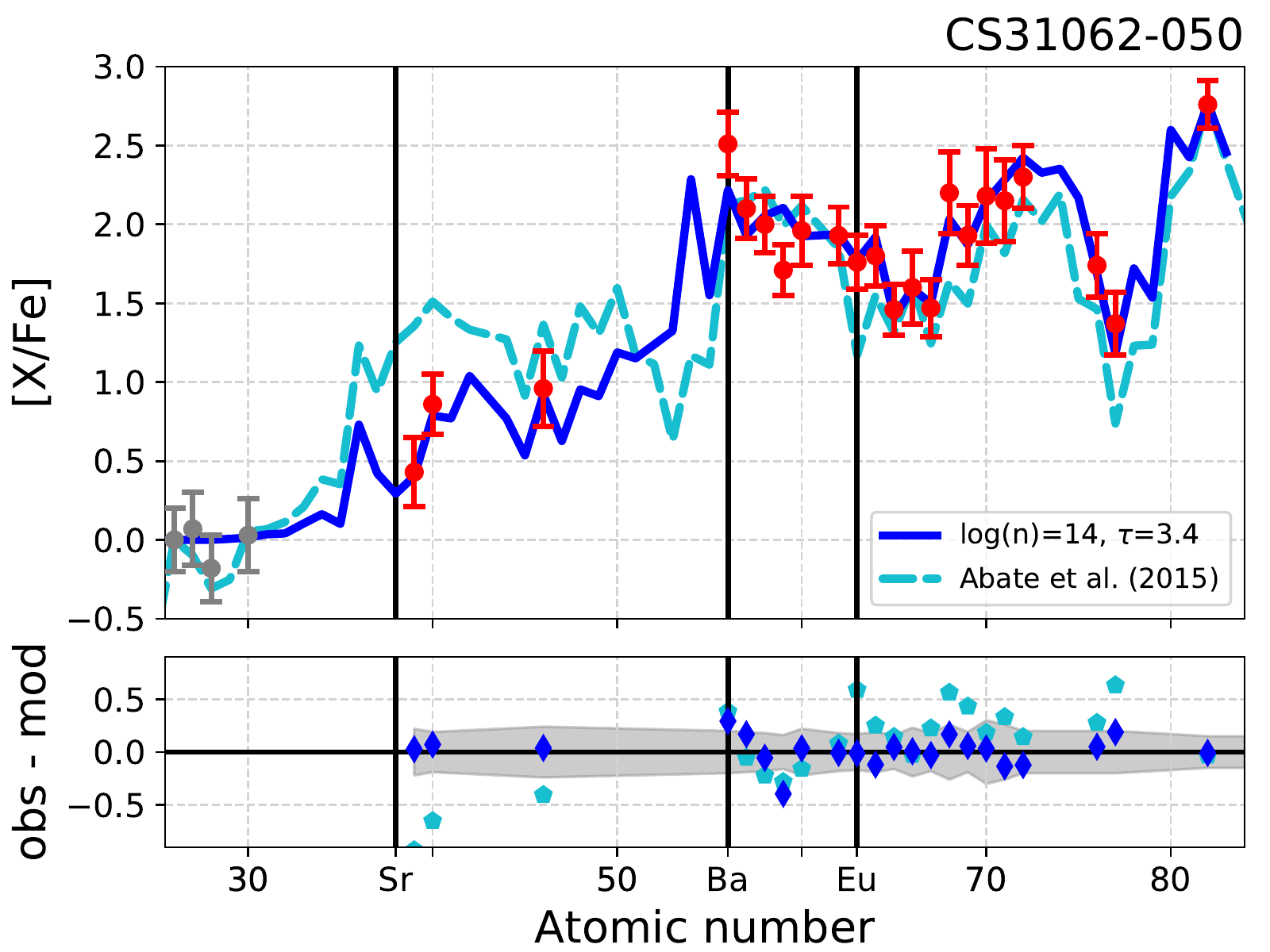}
\caption{Best fits of the heavy-element abundance pattern of CEMP-i star \CS. The blue line shows the best fitting i-process model with \n{14}. For comparison the best fit of \citet{Abate2015b} is shown in cyan. The best fitting s-process models with initial r-process enhancement can be found in Fig. 26 of \citet{Bisterzo2012}. The observational data are adapted from \citet{Johnson2004} with Ba, Os, and Ir abundances from \citet{Aoki2006}.}
\label{fig:CSstar}
\end{figure}

Figure \ref{fig:CSstar} shows the abundance pattern of CEMP-i star \CS, our best-fitting model, and the best s-process fit from \citet{Abate2015b}. \CS has a metallicity of \metal{-2.4} and it has been the subject of multiple observational studies \citep[e.g.][]{Aoki2002b, Aoki2002a, Lai2004, Johnson2004, Aoki2006, Lai2007} as well as i-process nucleosynthesis studies \citep{Dardelet2014, Hampel2016, Denissenkov2018}. The most comprehensive heavy-element abundance pattern is provided by \citet{Johnson2004} with measurements of 22 neutron-capture elements. \citet{Aoki2006} complemented this study by revising the uncertain abundance of Ba and deriving additional abundances for the interesting transition metals Os \Z{76} and Ir \Z{77}. 

Due to the high number of observed elements in \CS this star displays even more characteristics incompatible with s-process nucleosynthesis than already discussed for \LP, many of which have been pointed out by \citet{Johnson2004} and \citet[][see their Figure 26]{Bisterzo2012} but were not explicitly discussed in \pone. This includes a high abundance ratio of Eu and Tb \Z{65}, where both of these elements are predominantly produced in the Solar System by the r process with a contribution of 98\% and 94\%, respectively \citep{Sneden2008}. However, the observed ratio of $\mathrm{Eu}/\mathrm{Tb} = 3.1$ cannot be reproduced by a pure r process ($\mathrm{Eu}/\mathrm{Tb} = 1.52$), a pure s process ($\mathrm{Eu}/\mathrm{Tb} = 0.63$), or any combination thereof. Contrary to these two standard neutron-capture processes, an i process with \n{14} yields $\mathrm{Eu}/\mathrm{Tb} = 3.5$ and can naturally explain this high observed abundance ratio. Similar discrepancies for Dy \Z{66} and Ho \Z{67} can also be resolved by the i process. Additionally, abundances for Pd, Lu, Os and Ir (atomic numbers $Z=46$, 71, 76, and 77, respectively) are available for \CS. Due to the observational challenges posed by the spectral lines of these elements (\citealp[see, e.g., both Figures 1 of][]{Johnson2004} and of \citealp{Aoki2006}), they are infrequently studied in other CEMP stars. These rarely-measured elements provide valuable constraints on the heavy-element pattern between the three s-process peaks, all of which are well matched by our i-process models.

The best matching simulation for \CS requires a neutron exposure of \expo{3.4} at a density of \n{14} corresponding to a time of 2.5 days. The abundance pattern of \CS has also been the object of other i-process studies:  
\citet{Dardelet2014} reproduced the abundance pattern of \CS with single-zone i-process nucleosynthesis calculations at a neutron density of $n=3\times 10^{14}\,\mathrm{cm}^{-3}$ and \expo{40} (see their Figure 1). In comparison to our best-fitting model the neutron densities are of similar magnitude, while the required neutron exposures differ by more than an order of magnitude. With \expo{40}, our models overpredict the Pb abundance by at least 1.3 dex. However, it is not clear whether the observed Pb abundance was taken into account in the fits from \citet{Dardelet2014} and is reproduced by their high-exposure simulation. \citet{Denissenkov2018} show that the heavy-element abundances, including Pb, can be reproduced by the i process in their models of rapidly-accreting white dwarfs (see their Figure 12).

We find similar results for the neutron densities and exposures across the whole sample of studied CEMP-i stars. Table \ref{tab:fit_summary} lists the details of the best-fitting models for each of the stars and Appendix \ref{app:cemp_fits} (online only) shows the individual abundance patterns in detail. For all of the 24 studied stars (16 revisited from \pone), we find that the determined constant neutron density is either smaller than that found in \pone (for 9 stars) or the same (for 7 stars). The majority of 13 stars can be fitted best by an abundance pattern produced by a neutron density of \n{13}, five are best fitted by \n{14}, and only one by \n{15}. Despite \n{15} being the neutron density that is typically associated with the i process in the literature, we can rule out this neutron density as being characteristic for the CEMP-i stars.

\begin{figure}[t!bp]
\centering
\includegraphics[ width=0.49\textwidth]{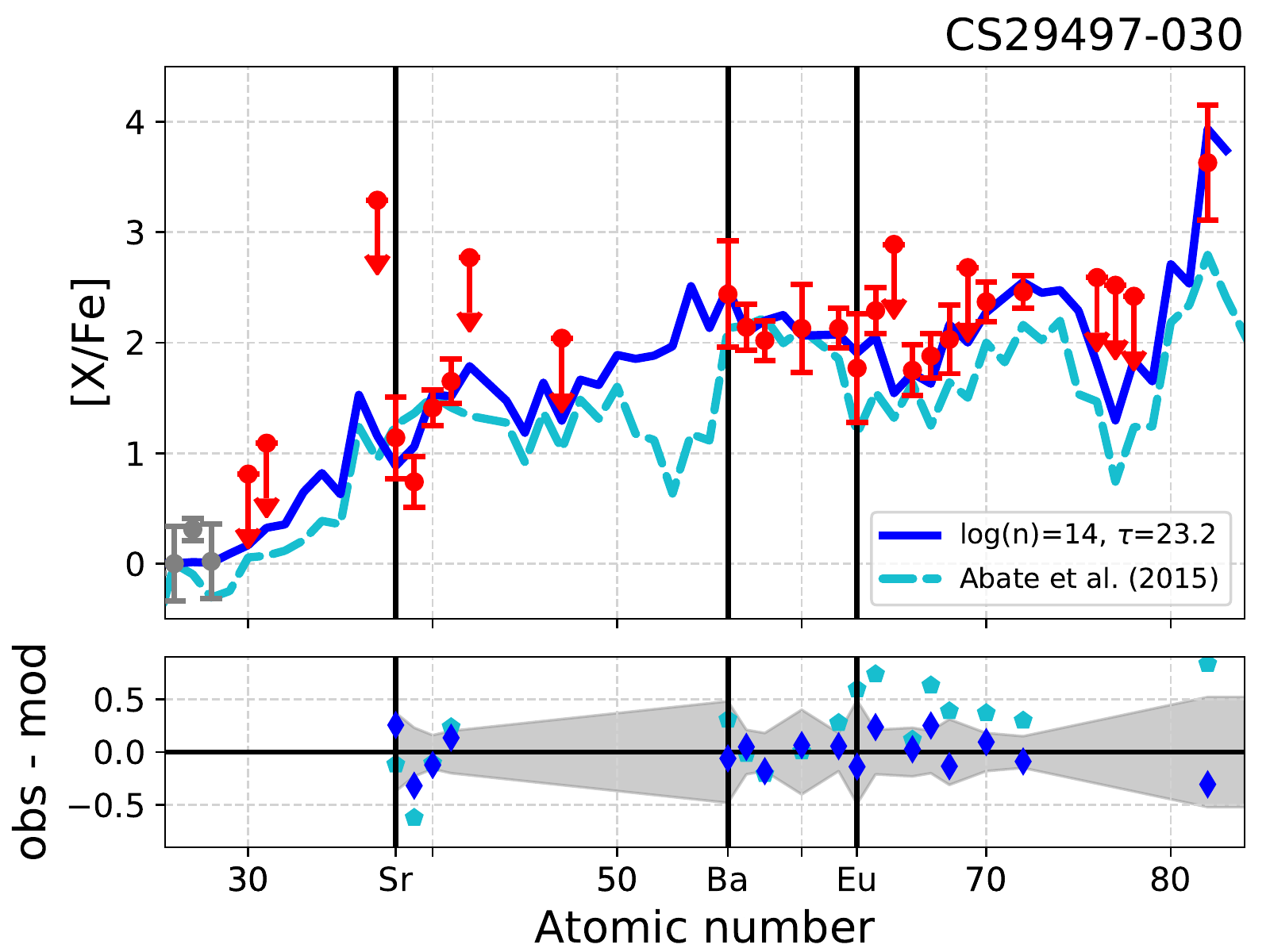} 
\caption{Best fits of the heavy-element abundance pattern of CEMP-i star \CShightau. The blue line shows the best fitting i-process model with \n{14}. For comparison the best fit of \citet{Abate2015b} is shown in cyan. The best fitting s-process models with initial r-process enhancement can be found in Fig. 18 of \citet{Bisterzo2012}.}
\label{fig:CS_high_tau}
\end{figure}

The integrated neutron exposures of the best-fitting models span a narrow range between $1.8 \leq \tau \leq 3.4 \,\, \mathrm{mbarn} ^{-1}$ with 4 exceptions:  

\begin{itemize}
\item CS29497-030 and CS22887-048 require much higher exposures of \expo{23.2} and \expo{12.0}, 
respectively,
\item HE2258-6358 requires a high neutron exposure of \expo{7.7} but is also the only star with a best fit at \n{15}, and
\item HE0336+0113 can only be matched with a model of a particularly low exposure of \expo{1.1}. 
\end{itemize}
CS29497-030 and CS22887-048 are the two stars with the highest Pb abundance amongst the studied CEMP-i stars ($\left[ \mathrm{Pb} / \mathrm{Fe} \right] = 3.5$ for both), as well as the highest ratios between the Pb and hs peak. Consequently, high neutron exposures are needed to reproduce these high Pb abundances. As an example, Figure \ref{fig:CS_high_tau} shows the abundance pattern of \CShightau with the best fitting models from this work and from \citet{Abate2015b}.

\begin{figure}[t!bp]
\centering
\includegraphics[ width=0.49\textwidth]{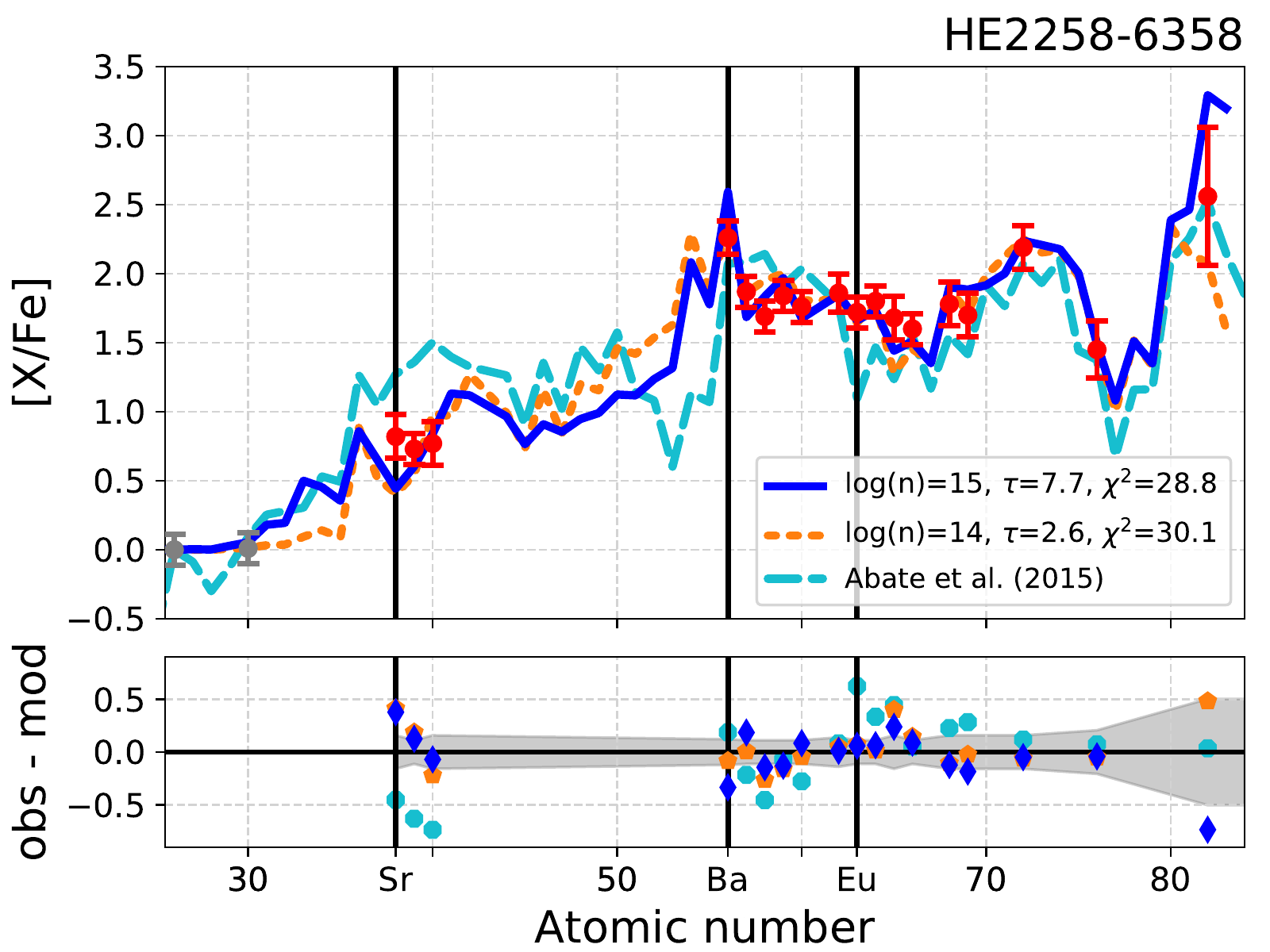} 
\caption{Best fits of the heavy-element abundance pattern of CEMP-i star \CShightau. The blue and orange lines shows the best fitting i-process models with \n{15} and \n{14}, respectively. For comparison the best fit of \citet{Abate2015b} is shown in cyan.}
\label{fig:HE_15}
\end{figure}

In contrast to CS29497-030 and CS22887-048, the abundance pattern of HE2258-6358, shown in Figure \ref{fig:HE_15}, does not show an extreme over abundance of Pb, which would naturally require a higher neutron expose. Instead, the exceptionally high required neutron exposure is due to the fact that the best fitting model has a neutron density of \n{15}. As evident from Figure \ref{fig:Pb_Ba_Sr_production} and discussed in \S\ref{sec:nuc-net-results}, the production of heavy elements needs higher exposures the higher the neutron densities. Figure \ref{fig:HE_15} also shows the second best fit to the abundances of HE2258-6358, which occurs at the parameters \n{14} and \expo{2.6}. The $\chi^2$ values of $28.8$ and $30.1$ for \n{15} and \n{14}, respectively, show that the quality of the fits are comparable and that the parameters of the alternative fit lie within the expected range.

\begin{figure}[t!bh]
\centering
\includegraphics[ width=0.49\textwidth]{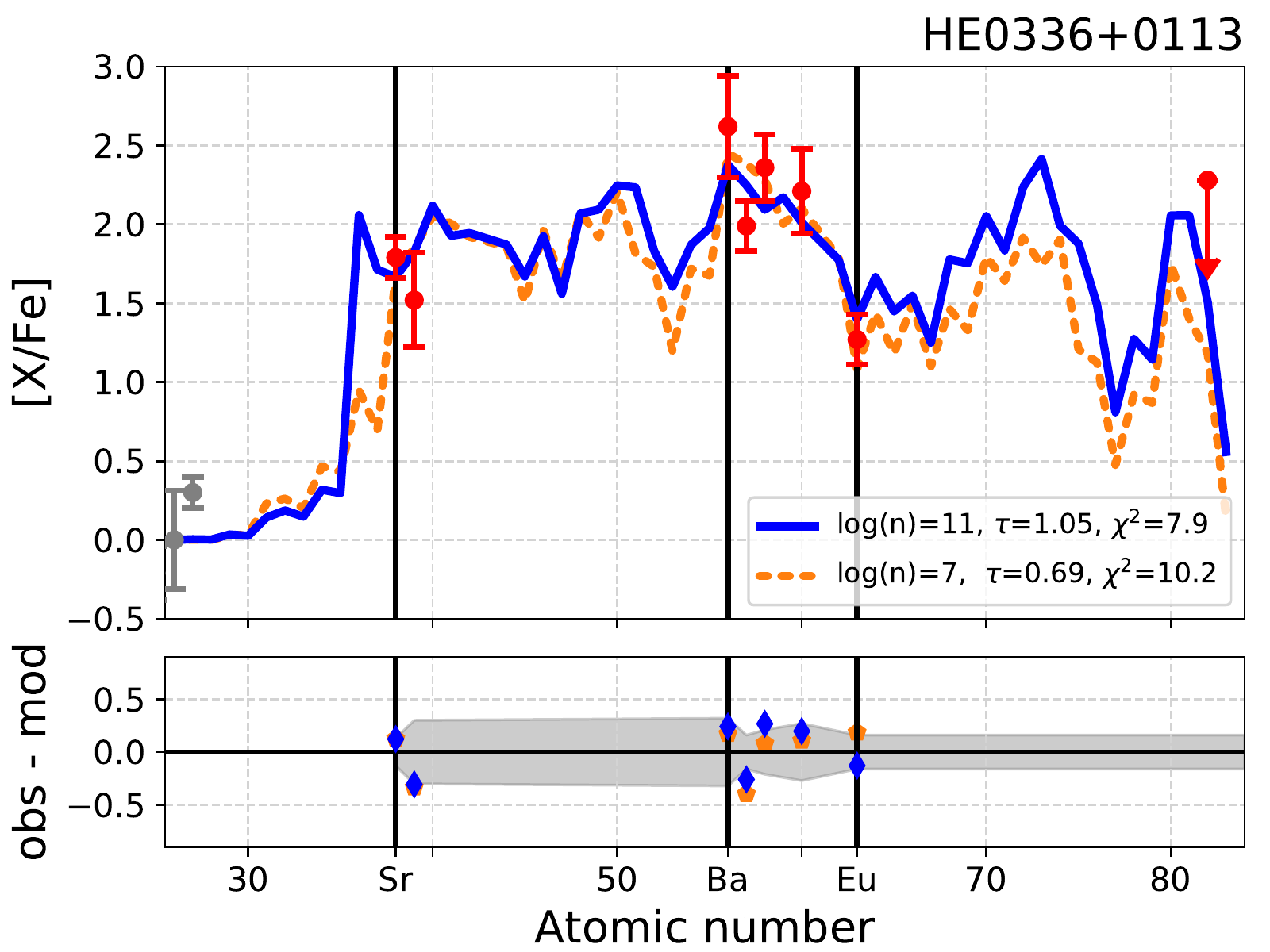} 
\caption{Best fits of the heavy-element abundance pattern of \HElowtau. The blue and orange line show the best fitting models with \n{11} and \n{7}, respectively. The best fitting s-process models with initial r-process enhancement can be found in Fig. 2 of \citet{Bisterzo2012}.}
\label{fig:CEMP_low_tau}
\end{figure}

The CEMP-i star whose fit to its abundance pattern results in the lowest neutron exposure is HE0336+0113, shown in Figure \ref{fig:CEMP_low_tau}. Not only does this fit indicate a low neutron exposure of \expo{1.1} but also a low neutron density of \n{11}.
 These parameters are very similar to those needed to match the post-AGB stars discussed in the next section, although the atmospheric parameters of $T = 5700 \pm 100 \, \mathrm{K}$ and $\log g = 3.5 \pm 0.25$  identify it as a subgiant and not a post-AGB star. In fact, the abundance pattern of HE0336+0113 has more similarities with those of the post-AGB stars than with the other CEMP-i stars: it has the highest ls peak amongst the CEMP-i stars ($\left[ \mathrm{Sr} / \mathrm{Fe} \right] = 1.8$) and a low [hs/ls] ratio combined with an upper limit of the Pb abundance at a particularly low level, which makes it the only CEMP star in the sample with $\sqbr{Pb}{Ba} < 0$. Moreover, it has a high overabundance of Ba compared to Eu of $\left[ \mathrm{Ba} / \mathrm{Eu} \right] = 1.3$. In addition to these properties, which are more characteristic for s- than for i-process nucleosynthesis, \citet[][Figure 2]{Bisterzo2012} matched the abundance pattern of HE0336+0113 by s-process nucleosynthesis of an AGB star with initial mass of 1.4\Msun and an initial r-process enrichment of $\left[ r / \mathrm{Fe} \right] = 0.5$.
For comparison, Figure \ref{fig:CEMP_low_tau} also shows the best fit to our simulation with a neutron density of \n{7}, which is typical for the s process. The resulting fit is not significantly worse than the one \n{11} and shows that determinations of the abundances of further elements, particularly of the heavy rare-earth elements, are desired to better constrain the nucleosynthetic history.

Our method for classifying CEMP-i stars did not have an upper limit for \sqbr{Ba}{Eu} ratios, although it is often used as criterion for s-process enrichment (see \S\ref{sec:stars}). In different classification schemes, e.g. $\left[ \mathrm{Ba} / \mathrm{Eu} \right] < 0.5$ \citep[e.g.][]{Beers2005}, HE0336+0113 would not be classified as a CEMP-i star, but as a CEMP-s star instead. However, the same classification scheme would also classify 12 other stars from our CEMP-i sample as CEMP-s, although our fits from \pone and  Table \ref{tab:fit_summary} show that their abundance patterns can be better matched by i-process nucleosynthesis. For example, this includes the previously discussed stars \LP (Figure \ref{fig:LPstar}) with $\sqbr{Ba}{Eu} = 0.9$ and \CS with $\sqbr{Ba}{Eu} = 0.7$ which have heavy-element abundances clearly incompatible with the s process and reproducible by i-process nucleosynthesis.

Although the limit of $\left[ \mathrm{Ba} / \mathrm{Eu} \right] < 0.5$ does not accurately distinguish CEMP-i and CEMP-s stars, the example of HE0336+0113 might indicate that this highest ratio of $\sqbr{Ba}{Eu} = 1.3$ could indeed be seen as an s-process indicator. Comparison with Figure 1 of \citet{Masseron2010} emphasises this outlier status of HE0336+0113 in the Ba-Eu-abundance plane in comparison to other CEMP-i stars: with $\sqbr{Ba}{Fe} = 2.6$ and $\sqbr{Eu}{Fe} = 1.3$ it is the only CEMP-i star that lies above the dashed lines which correspond to pure s-process predictions from metal-poor AGB stars, the highest of which lies at $\sqbr{Ba}{Eu} = 1.1$ \citep[see][]{Masseron2010}.

\begin{table*}[htbp]
  \centering
      \caption{Fit parameters for each star: number of observed elements $N_{\text{obs}}$ the fit is based on, neutron density $n$ in cm$^{-3}$ as found in this work and in \pone for comparison, time $t$ in seconds, integrated neutron exposure $\tau$ in mbarn$^{-1}$, dilution parameter $d$, and minimum $\chi^2$.}
     \begin{tabular}{cccccccc}
    \hline
        ID & $N_{\text{obs}}$ & $\log \left( n\right) $ & $\log \left( n\right) $ & $t$ & $\tau$ & $d$ & $\chi^2$ \\
        & & this work & \pone & (s) & (mbarn$^{-1}$) & & \\
\hline
\textbf{CEMP-i stars} & & & & & & & \\
BS16080-175 & 7 & 11 & 12 & $1.2 \times 10^{8}$ & 1.9 & 0.996 & 39.4 \\
BS17436-058 & 4 & 13 & 13 & $1.5 \times 10^{6}$ & 2.4 & 0.996 & 0.6 \\
CS22183-015 & 9 & 13 &  & $1.7 \times 10^{6}$ & 2.7 & 0.999 & 3.7 \\
CS22887-048 & 7 & 13 &  & $7.6 \times 10^{6}$ & 12.0 & 0.983 & 4.1 \\
CS22898-027 & 12 & 13 & 14 & $1.5 \times 10^{6}$ & 2.4 & 0.993 & 2.1 \\
CS22948-027 & 10 & 13 & 13 & $1.3 \times 10^{6}$ & 2.0 & 0.996 & 10.8 \\
CS29497-030 & 17 & 14 & 14 & $1.5 \times 10^{6}$ & 23.2 & 0.998 & 8.7 \\
CS29497-034 & 10 & 12 &  & $1.2 \times 10^{7}$ & 1.9 & 0.999 & 30.3 \\
CS29526-110 & 8 & 13 & 14 & $2.2 \times 10^{6}$ & 3.4 & 0.993 & 3.9 \\
CS31062-012 & 8 & 14 & 14 & $1.9 \times 10^{5}$ & 3.0 & 0.999 & 1.3 \\
CS31062-050 & 22 & 14 & 15 & $2.1 \times 10^{5}$ & 3.4 & 0.997 & 11.9 \\
HD187861 & 9 & 11 & 12 & $1.6 \times 10^{8}$ & 2.5 & 0.996 & 2.2 \\
HD209621 & 16 & 13 &  & $1.2 \times 10^{6}$ & 1.8 & 0.994 & 21.8 \\
HD224959 & 9 & 13 & 13 & $1.9 \times 10^{6}$ & 3.0 & 0.992 & 1.1 \\
HE0143-0441 & 9 & 12 & 14 & $1.7 \times 10^{7}$ & 2.7 & 0.992 & 5.9 \\
HE0243-3044 & 14 & 13 &  & $1.7 \times 10^{6}$ & 2.7 & 0.997 & 22.6 \\
HE0336+0113 & 7 & 11 &  & $6.7 \times 10^{7}$ & 1.1 & 0.997 & 7.91 \\
HE0338-3945 & 17 & 13 & 14 & $1.5 \times 10^{6}$ & 2.4 & 0.995 & 10.5 \\
HE0414-0343 & 14 & 14 &  & $1.9 \times 10^{5}$ & 3.0 & 0.999 & 12.0 \\
HE1305+0007 & 11 & 14 & 14 & $1.5 \times 10^{5}$ & 2.4 & 0.980 & 6.6 \\
HE1405-0822 & 19 & 13 &  & $1.5 \times 10^{6}$ & 2.4 & 0.999 & 54.4 \\
HE2148-1247 & 13 & 13 & 14 & $1.5 \times 10^{6}$ & 2.4 & 0.993 & 6.2 \\
HE2258-6358 & 18 & 15 & 14 & $4.9 \times 10^{4}$ & 7.7 & 0.998 & 28.8 \\
LP625-44 & 17 & 13 & 14 & $1.3 \times 10^{6}$ & 2.0 & 0.997 & 4.2 \\
 & & & & & & & \\
\textbf{post-AGB stars} & & & & & & & \\
IRAS07134 & 14 & 11 &  & $6.3 \times 10^{7}$ & 1.0 & 0.984 & 7.7 \\
IRAS22272 & 15 & 11 &  & $7.2 \times 10^{7}$ & 1.1 & 0.979 & 35.7 \\
J004441 & 15 & 11 &  & $7.6 \times 10^{7}$ & 1.2 & 0.978 & 6.6 \\
J050632 & 11 & 11 &  & $6.0 \times 10^{7}$ & 0.9 & 0.995 & 16.1 \\
J051848 & 14 & 12 &  & $8.2 \times 10^{6}$ & 1.3 & 0.982 & 33.9 \\
J052043 & 14 & 12 &  & $7.3 \times 10^{6}$ & 1.1 & 0.985 & 22.1 \\
J053250 & 12 & 12 &  & $8.2 \times 10^{6}$ & 1.3 & 0.993 & 9.5 \\

\hline
    \end{tabular}%
  \label{tab:fit_summary}%
\end{table*}%

\subsection{Comparison to post-AGB stars} \label{sec:pAGB-results}

\begin{figure}[t!bp]
\centering
\includegraphics[ width=0.49\textwidth]{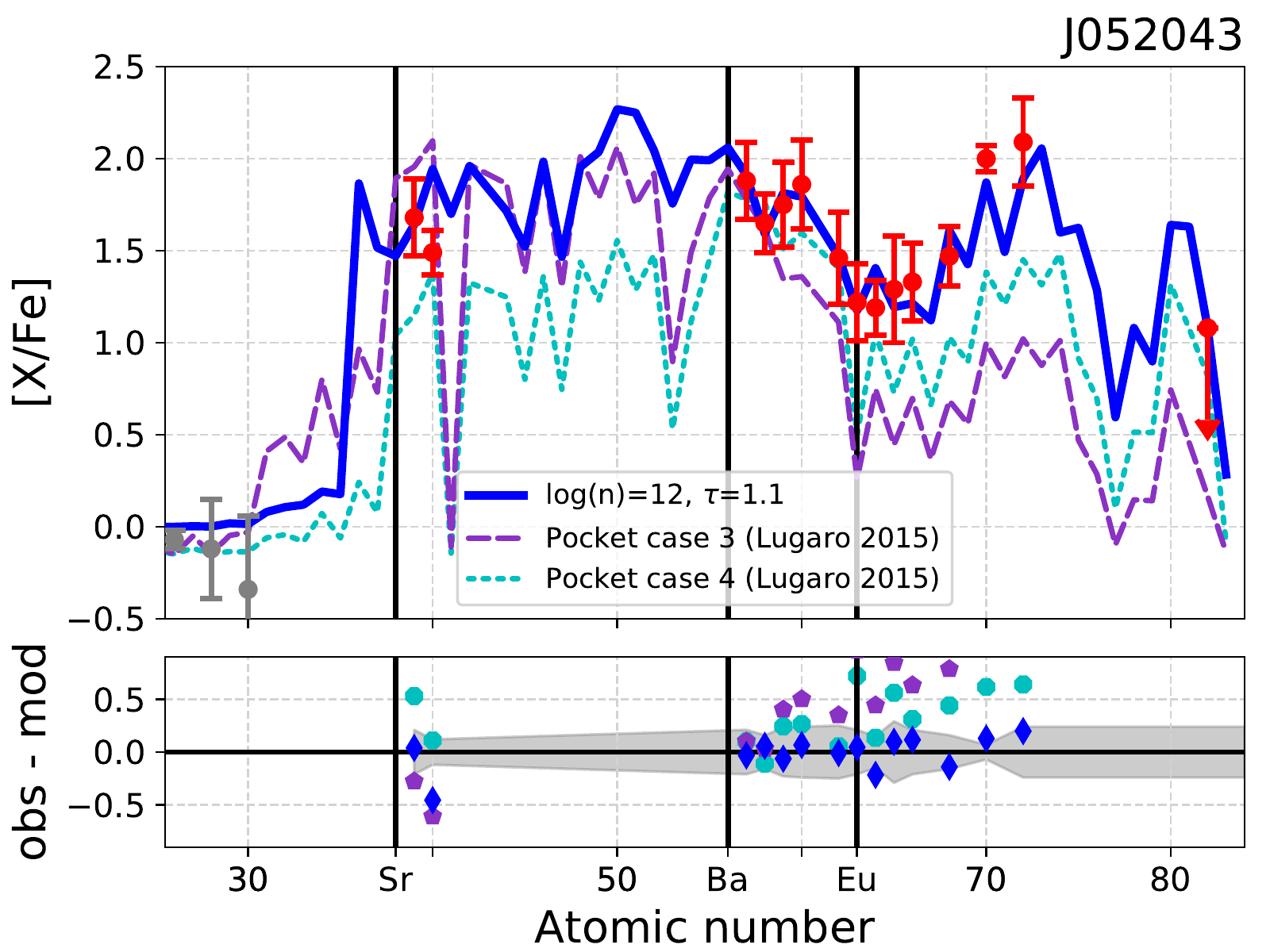}
\caption{Best fits of the heavy-element abundance pattern of post-AGB star \Jstar. The blue line shows the best fitting model with \n{12}. For comparison, the fits of \citet{Lugaro2015} are shown for modified \el{13}{C} pockets with two different proton fractions (normalised to La).}
\label{fig:pAGB}
\end{figure}

Figure \ref{fig:pAGB} shows the abundance pattern of the Pb-deficient post-AGB star \Jstar with our best-fitting model, while the fits for all other stars can be found in Appendix \ref{app:pAGB_fits} (online only). The best-fitting model was not only chosen by the minimal $\chi^2$ value of the fit but also by the additional constraint that the observed upper limit of the Pb abundance should not be overproduced. The abundance pattern of \Jstar can be best reproduced by simulations with a neutron density of \n{12} for 83.3 days, resulting in a neutron exposure of \expo{1.1}.

For comparison, s-process models from \citet{Lugaro2015} are also shown in Figure \ref{fig:pAGB}. These are the results of AGB evolution and nucleosynthesis of models with an initial mass of 1.3 $M_{\odot}$, metallicity \metal{-1.3} and \el{13}{C} pockets produced by two different constant proton abundances mixed into the intershell with ${X \left( \mathrm{H} \right)_{\mathrm{case 3}} = 0.7 \times 10^{-4}}$ and $X \left( \mathrm{H} \right)_{\mathrm{case 4}} = 1.05 \times 10^{-4}$. These two different \el{13}{C} pockets were fine-tuned to comply with the upper Pb limit and to match the hs abundances, but fail to reproduce the other characteristics of the observed abundance patterns. This can be clearly seen by the underproduction of the elements between the hs and Pb peak starting at Eu. The enhancement levels of these elements can be naturally matched by our simulation without overproducing Pb. Additionally, our simulations are successful at reproducing the observed ls- and hs-enhancements and the relative strengths of the s-process peaks.

The parameters for the best fitting i-process models for the studied post-AGB stars are summarised in Table \ref{tab:fit_summary}. The abundance patterns and upper Pb limits of all 7 post-AGB stars can be fitted best by i-process models of \n{11} and \n{12}, which is at the lower end of the tested neutron densities. Similar to the CEMP-i stars, our models show that the process responsible must have operated on short time scales. The required integrated neutron exposures for the post-AGB stars lie in the narrow range between $1.0 \leq \tau \leq 1.3 \, \mathrm{mbarn} ^{-1}$ leading to time scales between 0.2 and 2.5 years. Overall, the neutron densities and exposures found for the post-AGB stars are lower than those required to match the CEMP-i stars. The lower neutron densities in the i-process simulations allow for somewhat longer operation time scales in the post-AGB stars compared to CEMP-i stars. In general, the time scales of all of the best-fitting models are by far shorter than those of typical s-process nucleosynthesis in the \el{13}{C} pocket of the order of $10^4 \,$yr: the longest of our simulations fitting the post-AGB stars requires only 2.4 years.

\subsection{Summary of Results} \label{sec:result_summary}

\begin{figure}[t!bh]
\centering
\includegraphics[ width=0.49\textwidth]{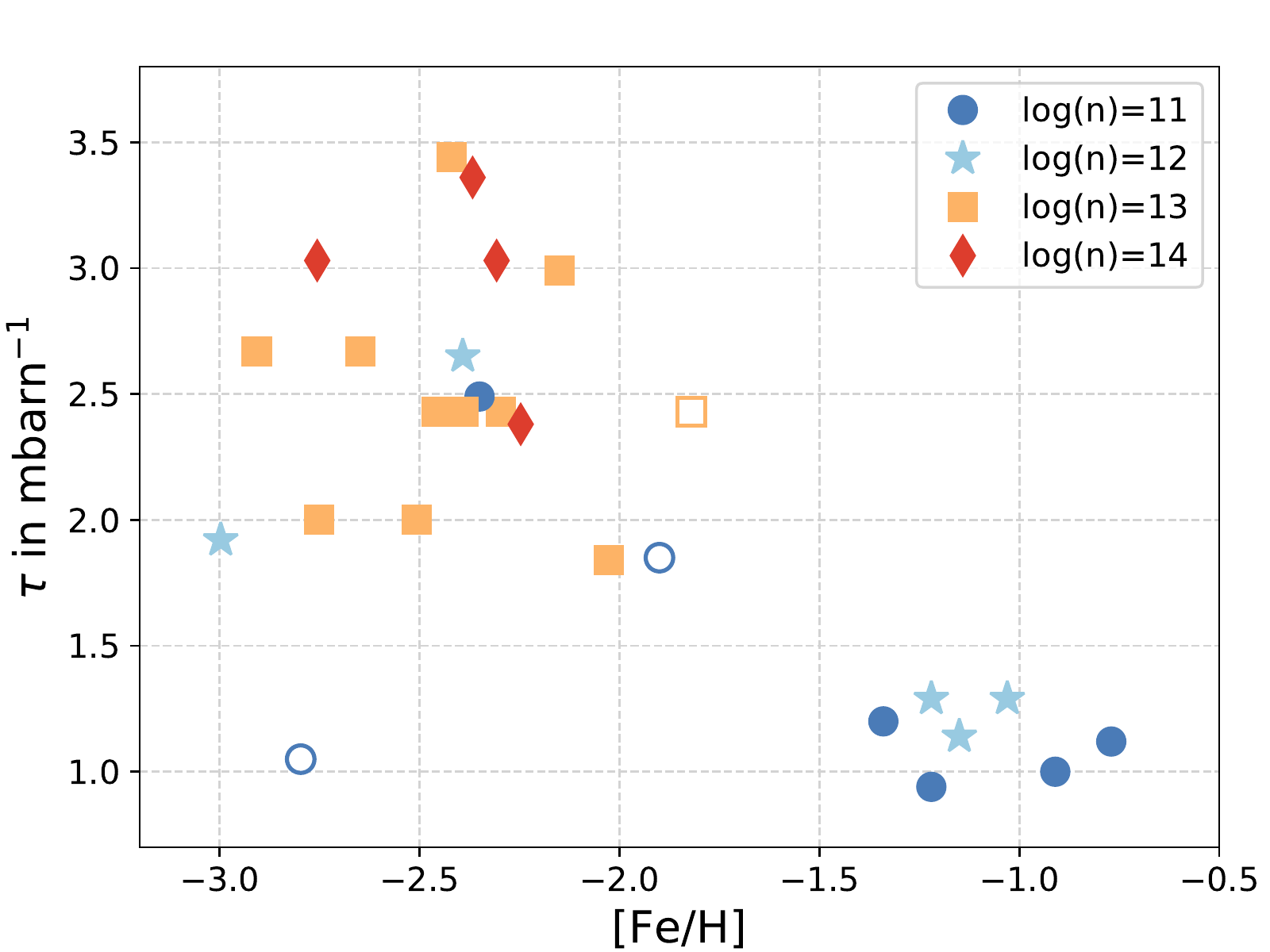} 
\caption{Neutron density and exposure of the best fitting models for each star as function of the metallicity. The empty symbols denote the least reliable fits due to only 7 or less measured heavy-element abundances. Note that only exposures up to \expo{3.5} are shown and the three CEMP-i stars with higher exposures are not included in this figure. See discussion in main text for further details, also about the outlying CEMP star \HElowtau at \metal{-2.8} and \expo{1.1}. }
\label{fig:tau_FeH}
\end{figure}

Figure \ref{fig:tau_FeH} shows the exposure and neutron density of the best fits to the observations as a function of their metallicity.
Interestingly, our results show a bimodial distribution of parameters characterising the best-matching i-process conditions. The CEMP-i fits predominantly have higher neutron densities of \n{13} to \n{14} and higher integrated neutron exposures of $\tau \geq 1.8\,\mathrm{mbarn}^{-1}$ than the simulations matching the abundances of the Pb-poor post-AGB stars, which show lower neutron densities of \n{11} to \n{12} and lower integrated neutron exposures of $1.0 \leq \tau \leq 1.3\,\mathrm{mbarn}^{-1}$. A possible explanation of this bimodality is that these are the products of i-process nucleosynthesis occurring at two different sites where the specific i-process conditions give rise to slightly different characteristics. Alternatively, we could assume the same i-process site and that the reduction of $n$ and $\tau$ is an effect of changing metallicity on the neutron production.

\section{Discussion} \label{sec:Discussion}

The two fundamental questions regarding the i process are: \textit{Does it happen in nature? And if so, where and how?}  
Despite the simplicity of our one-zone models, our results confirm that the observed abundance patterns of CEMP-i and Pb-poor post-AGB stars are best explained by i-process nucleosynthesis. This result can be used to help us shed light on possible i-process sites by considering which progenitors may be responsible for the observed i-process abundance patterns.

Let us start with the more obvious case: we know that the progenitors of the Pb-poor post-AGB stars are low-mass AGB stars and that their heavy-element enrichment is a direct result of their previous stellar evolution. Therefore our results imply that low-mass AGB stars with initial masses of approximately  $1 - 1.5 \Msun$ at metallicities of $-1.3 \leq \left[ \mathrm{Fe} / \mathrm{H} \right] \leq -0.7$ reach stages in their evolution that give rise to i-process conditions. Moreover, it seems that these progenitors have not experienced standard AGB s-process nucleosynthesis, as this would produce higher Pb abundances accompanying the other heavy-element enhancements \citep{De-Smedt2012, De-Smedt2014, Lugaro2015, Trippella2016}.
From our results we cannot constrain the evolutionary phase in which the i-process nucleosynthesis occurred. Did it occur during the TP-AGB phase, replacing the s process, or is it the result of late He-shell flashes when the star has moved to the post-AGB?

The high neutron densities required for the i process could result from proton-ingestion episodes (PIEs).
PIEs occur when protons are entrained into convective layers rich in He and C. This can lead to a neutron burst at i-process neutron densities via the $\el{12}{C}\left( \mathrm{p}, \gamma\right)\el{13}{N}\left( \beta^+ \nu \right)\el{13}{C}\left( \alpha, \mathrm{n}\right)\el{16}{O}$ reaction chain \citep{Cowan1977}. A diversity of sites has been proposed  to host PIEs and in extension i-process nucleosynthesis. These include core He flashes in low-metallicity low-mass stars \citep{Campbell2010, Cruz2013}, super-AGB stars \citep{Doherty2015, Jones2016}, very late thermal pulses \citep{Herwig2011}, rapidly-accreting white dwarfs \citep{Denissenkov2017, Denissenkov2018}, and metal-poor massive stars \citep{Clarkson2018, Banerjee2018}.
PIEs have also been predicted to occur in the first fully developed convective thermal pulse(s) in low-mass, low-metallicity AGB
stars where the upper metallicity limit depends on mass and composition but tends to lie below $\left[ \mathrm{Fe} / \mathrm{H} \right] \lesssim -2$ \citep{Hollowell1990, Fujimoto1990, Iwamoto2004, Campbell2008, Lau2009, Cristallo2009b, Cristallo2016}. While this metallicity is compatible with the studied CEMP-i stars, it is questionable whether this is also a realistic scenario to have occurred in the progenitors of the Pb-poor post-AGB stars of higher metallicity. 

How could the i process occur in a post-AGB star with  $\left[ \mathrm{Fe} / \mathrm{H} \right] > -2$? Sakurai's Object (V4334 Sagittarii) is a born-again giant at solar metallicity and shows neutron-capture enrichments but only in the lightest elements such as Rb, Sr, Y, and Zr \citep{Asplund1999}. Similarly to the post-AGB stars in our sample, Sakurai's Object does not show the abundances that are expected from s-process nucleosynthesis during the TP-AGB phase at the observed solar metallicity. In particular, the measured abundances of the s-process peaks relative to each other pose challenges to standard s-process nucleosynthesis with a significantly higher observed overproduction of the ls peak. Without significant s-process contribution during the TP-AGB phase, the observed
neutron-capture nucleosynthesis could be driven by a late He-shell flash which ingests the thin remaining H envelope. \citet{Herwig2011, Herwig2014} studied such a (very) late He-shell flash in three-dimensional hydrodynamic simulations. The resulting i-process nucleosynthesis informed by the three-dimensional calculations can successfully explain the abundances in Sakurai's Object \citep{Herwig2011}. 

The complicated details of different effects in the three-dimensional hydrodynamic simulations of proton-ingestion episodes are still debated and not fully understood. Stellar evolution models predict a prompt splitting of the convective zone prohibiting further mixing and i-process nucleosynthesis. Three-dimensional hydrodynamic simulations motivate that such a split can be delayed \citep{Herwig2011} whereas \citet{Stancliffe2011} did not find evidence for convective zone splitting. Additionally, \citet{Herwig2014} observed that the proton ingestion can trigger global oscillations with uncertain consequences, such as enhanced entrainment rates or self-quenching.

Different to the post-AGB stars, the nature of the progenitor that hosted the i-process nucleosynthesis observed in CEMP-i stars is less clear. The observed abundance patterns of CEMP-i stars are assumed to show the products of nucleosynthesis in a companion star that have been accreted onto the lower-mass secondary in a binary system that we observe today. Therefore we do not know the initial mass and evolution of the progenitor. However, based on arguments from population-synthesis and from comparing the relative numbers of CEMP-s and CEMP-i stars, the formation-channel of CEMP-i stars is not expected to be much less common than that of CEMP-s stars, if not even equally as likely \citep{Abate2016}.   

This scenario of pollution by a companion star is supported by the large binary fraction amongst the CEMP stars with Ba enrichment, which could be consistent with all these stars being in binaries \citep{Lucatello2005, Starkenburg2014}. However, \citet{Hansen2016} found a binary fraction of $\approx 80 \%$ in their sample of CEMP-s and CEMP-i stars using a systematic and precise long-term radial-velocity monitoring program. While this study confirms a much higher binary fraction amongst CEMP stars with Ba enrichment compared to normal metal-poor giants, it also finds four CEMP-s stars without any signs of a companion. A population-synthesis study by \citet{Abate2018} shows that at least some or even all four of these apparently single CEMP-s stars could be undetected binaries with orbital periods of $\mathrm{P} \gtrsim 10^4$ days. Alternatively, these stars could be single stars that require an alternative formation mechanism, e.g., pollution of their birth-clouds from previous-generation spinstars \citep{Choplin2017}.  
The mentioned studies are for CEMP stars with Ba enrichment in general, which include CEMP-s and CEMP-i stars. It will need larger samples of stars with consistent radial velocity monitoring in order to draw individual conclusions for CEMP-s and CEMP-i stars separately. Due to the limited data, the binary fraction of CEMP-i stars in particular remains unclear.

If the binary fraction of the CEMP-i stars is similar to that of CEMP-s and if the stars without radial-velocity variations from \citet{Hansen2015} are true single stars instead of undetected long-period binaries, then there may be the need for an i-process site that can account for a minority of single CEMP-i stars that are not polluted by a companion star.
For instance, metal-free and metal-poor massive stars of $20-30 \Msun$ are considered as candidates for polluting the interstellar medium with i-process material from which a single CEMP-i star could subsequently form \citep{Banerjee2018, Clarkson2018}.

Alternatively, \citet{Denissenkov2017, Denissenkov2018} investigated rapidly-accreting white dwarfs (RAWDs) as i-process sites and also as progenitors for single CEMP-i stars. Under certain conditions of stable and rapid mass transfer between white dwarfs in a multiple system, i-process nucleosynthesis could be hosted in the resulting He shell flashes. At decreasing metallicity, \citet{Denissenkov2018} find increasing mass retention efficiencies for RAWDs which opens a possible channel for supernova type Ia explosions. If such a RAWD system in a close binary is orbited by a companion in a wider tertiary, this companion could be polluted with i-process material from the RAWD. Moreover, if a subsequent supernova explosion leads to the ejection of this tertiary companion from the triple system, it could be observable today as a single CEMP-i star \citep{Denissenkov2018}.

The formation of CEMP-i stars by RAWDs requires a very specific sequence of events, which may be disfavoured by the stellar population. While the mechanism is possible, only population synthesis calculations can determine if it is probable. The i-process nucleosynthesis in RAWDs only results from stable burning of accreted H, for which a very narrow range of accretion rates is required \citep{Nomoto2007}. So far the stable mass transfer has been imposed in the simulations by \citet{Denissenkov2017, Denissenkov2018} but for low-metallicity systems with $\FeH < -1$ stable mass transfer at this specific rate is predicted to be unlikely \citep{Kobayashi1998}. 
Moreover, it is not clear yet whether RAWDs are viable progenitors of CEMP-i stars, due to the low occurance rates of triple systems, in particular of such specific configurations that ultimately can form a close CEMP-i binary or are able to eject a CEMP-i single star from the system \citep{Rastegaev2010, Hamers2013, De-Marco2017}.

When using our parametric study to infer i-process parameters that reproduce the observed abundance patterns, one has to keep the simplicity of our models in mind. We utilised single-zone nuclear network calculations to study i-process nucleosynthesis detached from an actual stellar framework. While we used representative temperatures and densities derived from stellar evolution models, we still keep them constant throughout our calculations. Most significantly, we even detach the production of the neutrons not only from the stellar site and the physical conditions, but also from the ongoing nucleosynthesis. While this has the advantage that we can concentrate on the details of the heavy-element production, it means that we only study the influence of a constant neutron density. 
Important aspects as for example the production of neutrons, its metallicity and time dependence, the effect of mixing and the replenishment of neutron-capture seeds etc. remain untouched in this study.

Additionally, the nuclear data and reaction rates at the base of this study suffer from significant uncertainties. It is known that their effects on i-process abundance patterns can be large: 
\citet{Bertolli2013} found, for example, that predicted i-process abundances of [Ba/La] can change up to 1 dex depending on the theoretical nuclear physics models and effects of up to 0.3 dex were found to come from the (n, $\gamma$) reaction rate uncertainties of ls elements \citep{Denissenkov2018b}.
Therefore, it is important to systematically study the influences that nuclear physics uncertainties have on the simulated abundance patterns.

\section{Conclusions} \label{sec:summary}

In this study we examined the observed heavy-element abundances of two types of objects that show enrichments in traditional s-process elements produced in AGB stars but whose abundance patterns are generally incompatible with s-process predictions: CEMP-i stars and Pb-poor post-AGB stars, with representative metallicities of $\left[ \mathrm{Fe} / \mathrm{H} \right] \approx -2.5$ and $\left[ \mathrm{Fe} / \mathrm{H} \right] \approx -1.3$, respectively. We can explain these abundance patterns, including the combination of puzzlingly low Pb enhancements and high rare-Earth element abundances, as results of i-process nucleosynthesis. We used nuclear-network calculations to study heavy-element production at different constant neutron densities up to \n{15}. The constraints posed by measured Pb abundances in these objects allowed us to characterise the neutron densities and exposures of the process responsible for creating the observed heavy-element abundances. 

We find that the patterns of the post-AGB stars are best explained by neutron-capture nucleosynthesis at relatively low neutron densities of \n{11} or \n{12} and exposures between \expo{1.0} and \expo{1.3}. In contrast, higher neutron exposures of at least $\tau > 2.0 \, \mathrm{mbarn}^{-1}$ are required to reproduce the abundance patterns of CEMP-i stars as well as higher neutron densities, mostly at \n{13} or \n{14}. These results offer new constraints and insights regarding the properties of i-process sites and demonstrate that the responsible process operates on short time scales, of the order of a few years or less, depending on the neutron density, and requires lower bulk neutron densities than the initially characteristic i-process neutron density of \n{15} proposed by \citet{Cowan1977}.  

That the $n$ and $\tau$ parameters of the best fits to the abundance patterns of CEMP-i stars and Pb-poor post-AGB stars cluster in different regions of the parameter space might offer additional insights: there could be a metallicity dependence of the underlying process that leads to the neutron bursts, or this could indicate two different mechanisms or sites. 

To understand which neutron density over which time scale is responsible for which abundance patterns is only a first step. Future work will need to study how representative these patterns are for more realistic neutron-density profiles, which are expected to reach some peak density and then decline afterwards. 
We still have to identify the main influences that shape the resulting abundance pattern and many questions remain unsolved: How does the lower neutron-density tail affect the final abundances after the peak density is reached? How quickly does the neutron source have to cease in order to maintain the characteristic i-process pattern formed at the peak neutron density before the exposure at lower neutron density reshapes the abundance patterns to resemble that of a typical s process? Can these effects (or the lack thereof) provide insights into the physical and structural properties of the i-process site?
In order to understand the mechanisms responsible for i-process conditions, future theoretical i-process studies need to simulate heavy-element production in more realistic stellar environments, for example as presented for RAWDs by \citet{Denissenkov2018}. A larger number of objects over a wider range of metallicities will improve the certainty with which we can constrain i-process properties and, of course, is fundamental to answering questions like \textit{How common is i-process nucleosynthesis?}
An increase in detected and (ideally homogeneously) analysed objects with i-process abundance patterns will therefore be very enriching for the field.

We are still at the beginning of understanding i-process nucleosynthesis and there are many open questions that require further theoretical and observational studies. Ultimately improving our understanding of the mechanisms leading to the conditions under which the i-process operates will help us to complete our understanding of stellar physics, and the evolution and the origin of heavy elements in the Galaxy. 




\acknowledgments
The authors thank Simon Campbell and Carlo Abate for very helpful discussion and the referee for their careful review and useful comments which have helped to improve this paper. A.I.K. acknowledges financial support from the Australian Research Council (DP170100521). This work is supported by the Lend\"ulet grant (LP17-2014) of the Hungarian Academy of Sciences to M.L. and the NKFIH KH\_18 (130405) project. B.S.M. gratefully acknowledges funding from NASA Grant No. NNX17AE32.






\bibliography{library}



\appendix

\section{All fits: CEMP-i stars (online only)} \label{app:cemp_fits}
\begin{figure}[h]
\begin{minipage}[b]{.45\textwidth}
\centering
\includegraphics[ width=\linewidth]{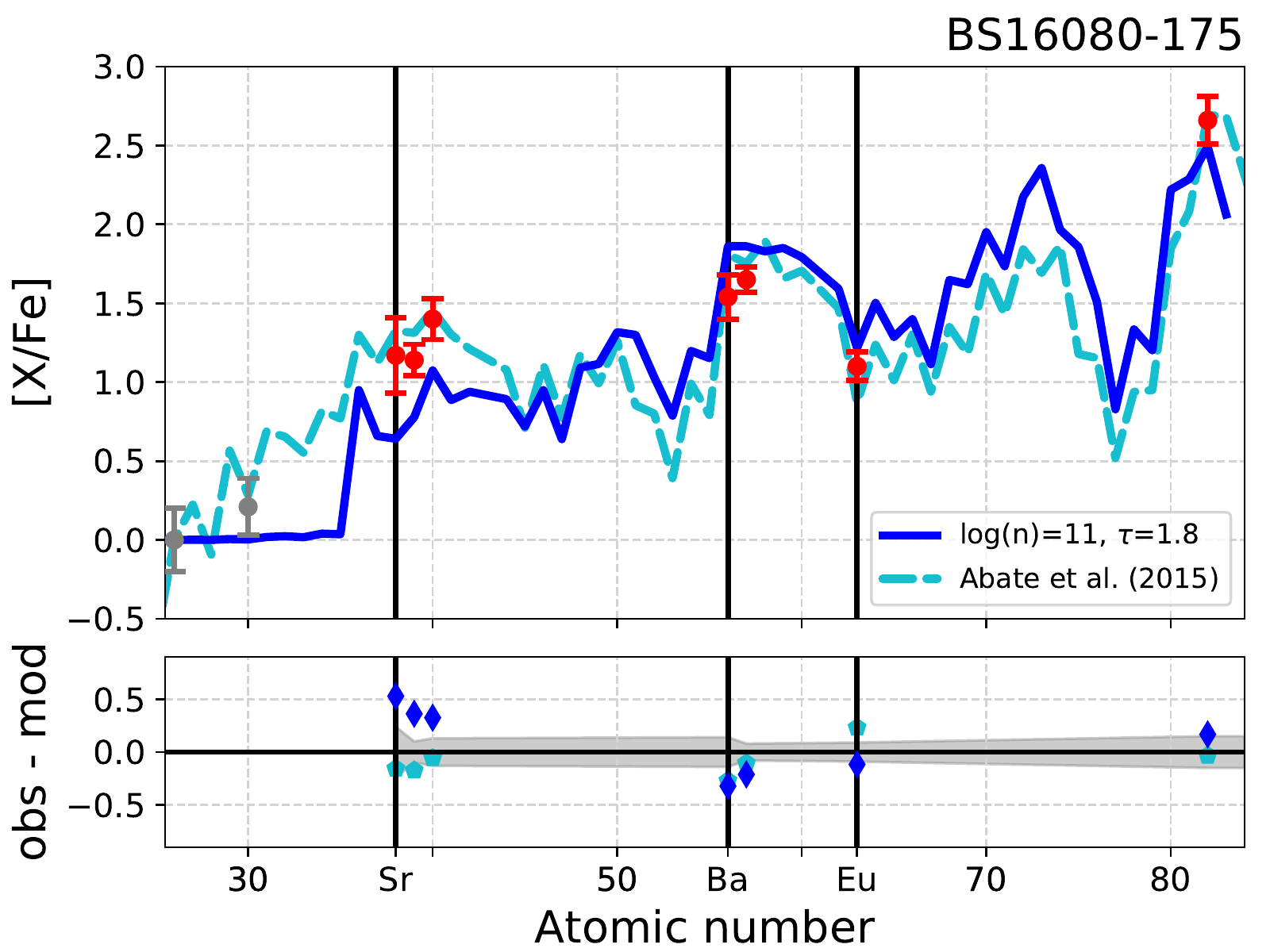}
\caption{Best fitting model for CEMP-i star BS16080-175 compared to the best fit of s-process nucleosynthesis with binary evolution of \citet{Abate2015b}. The best fitting s-process models with initial r-process enhancement can be found in Fig. 5 of \citet{Bisterzo2012}.} \label{Fig:app_cemp_first}
\end{minipage}
\hfill
\begin{minipage}[b]{.45\textwidth}
\centering
\includegraphics[ width=\linewidth]{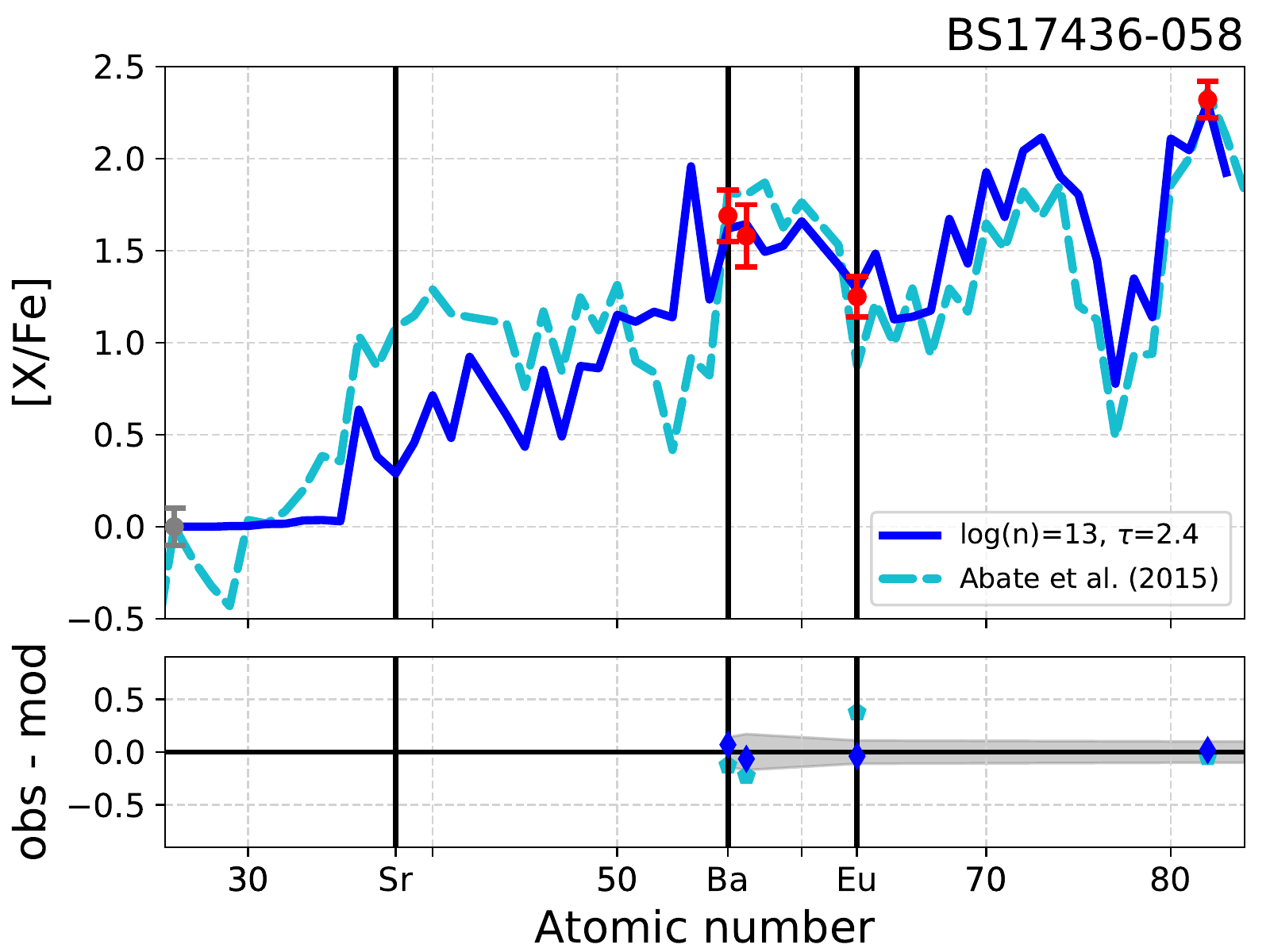}
\caption{Best fitting model for CEMP-i star BS17436-058 compared to the best fit of s-process nucleosynthesis with binary evolution of \citet{Abate2015b}. The best fitting s-process models with initial r-process enhancement can be found in Fig. 16 of \citet{Bisterzo2012}.}
\end{minipage}
\hfill
\begin{minipage}[b]{.45\textwidth}
\centering
\includegraphics[ width=\linewidth]{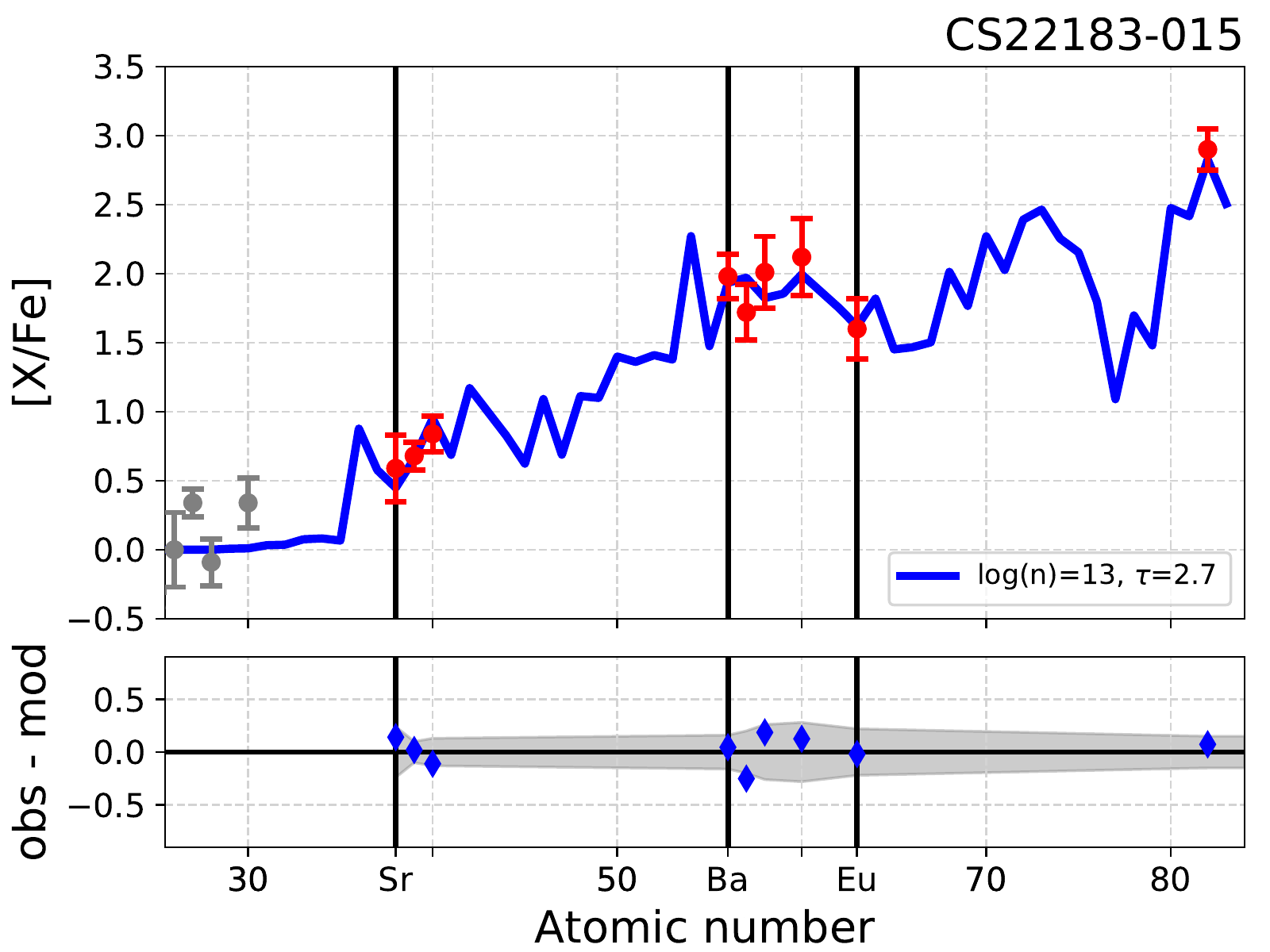}
\caption{Best fitting model for CEMP-i star CS22183-015. The best fitting s-process models with initial r-process enhancement can be found in Fig. 32 of \citet{Bisterzo2012}.}
\end{minipage}
\hfill
\begin{minipage}[b]{.45\textwidth}
\centering
\includegraphics[ width=\linewidth]{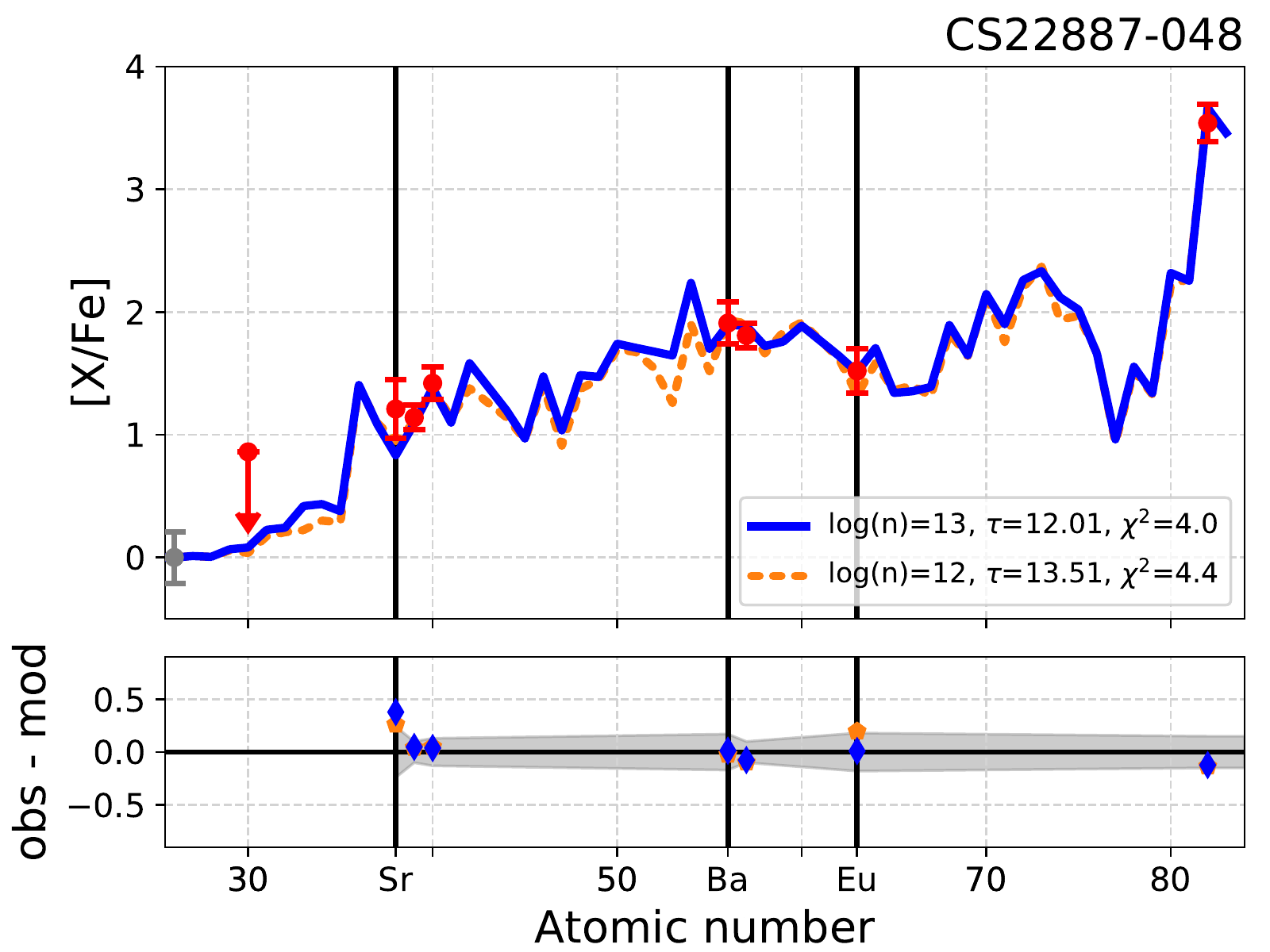}
\caption{Best fitting models for CEMP-i star CS22887-048. The best fitting s-process models with initial r-process enhancement can be found in Fig. 36 of \citet{Bisterzo2012}.}
\end{minipage}
\hfill
\begin{minipage}[b]{.45\textwidth}
\centering
\includegraphics[ width=\linewidth]{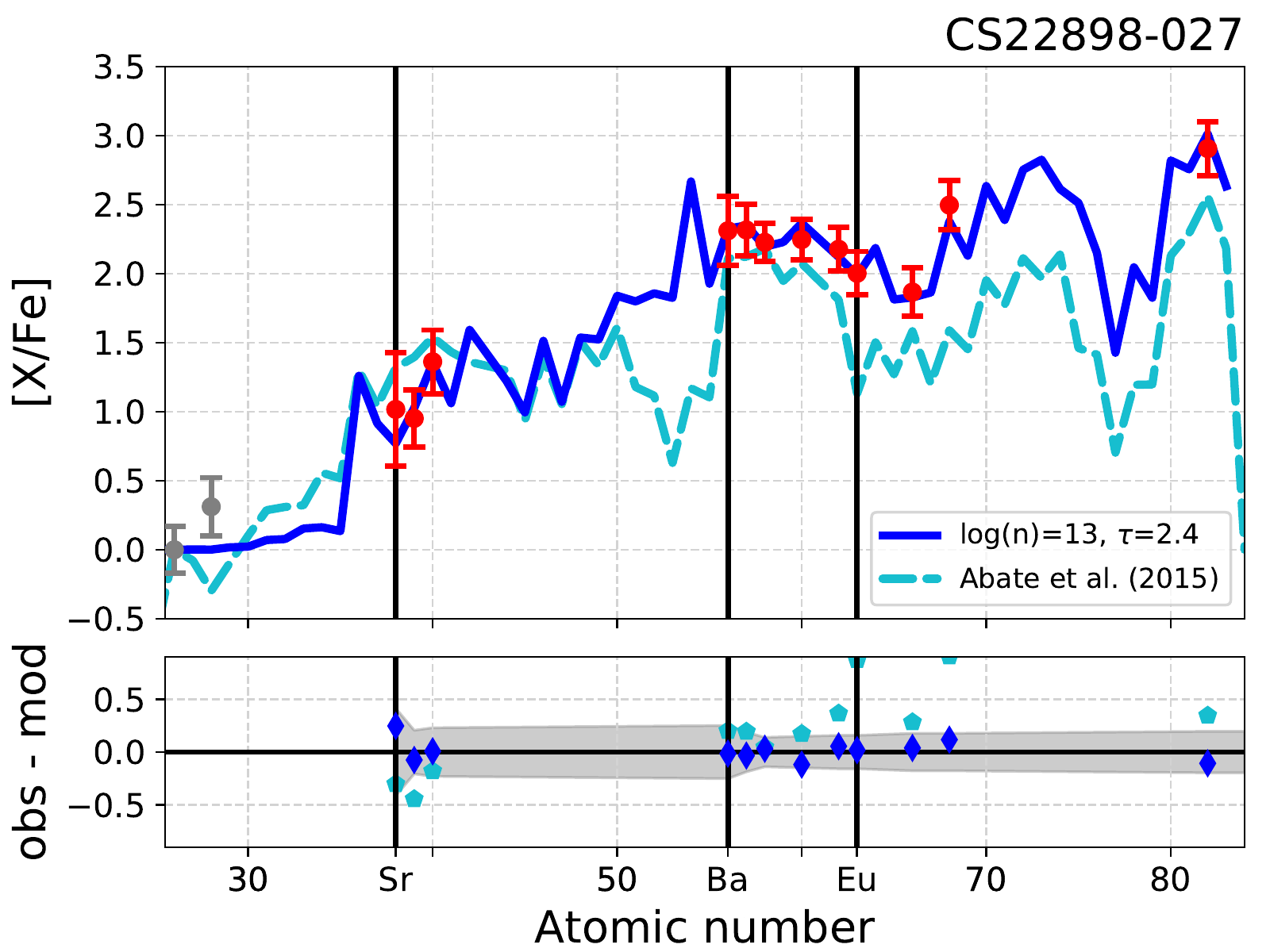}
\caption{Best fitting model for CEMP-i star CS22898-027 compared to the best fit of s-process nucleosynthesis with binary evolution of \citet{Abate2015b}. The best fitting s-process models with initial r-process enhancement can be found in Fig. 17 of \citet{Bisterzo2012}.}
\end{minipage}
\hfill
\begin{minipage}[b]{.45\textwidth}
\centering
\includegraphics[ width=\linewidth]{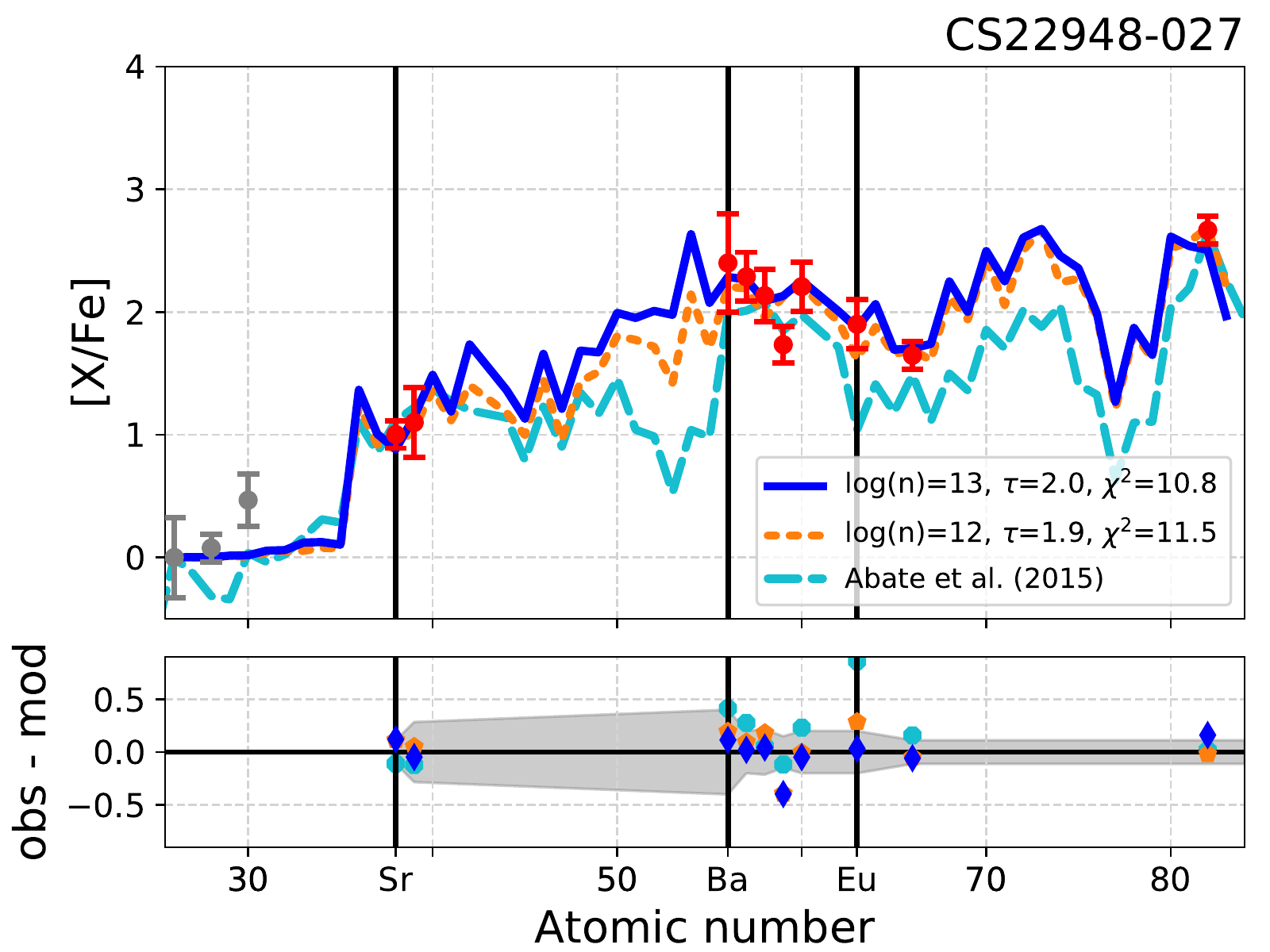}
\caption{Best fitting models for CEMP-i star CS22948-027 compared to the best fit of s-process nucleosynthesis with binary evolution of \citet{Abate2015b}. The best fitting s-process models with initial r-process enhancement can be found in Fig. 27 of \citet{Bisterzo2012}.}
\end{minipage}
\hfill
\end{figure}

This section shows the best fitting models for each of the CEMP-i stars that are not shown in section \S\ref{sec:CEMP-results} in comparison to the observed abundance patterns in Figures \ref{Fig:app_cemp_first} to \ref{Fig:app_cemp_last}.  
Details of each best fit (neutron
density $n$ and time-integrated neutron exposure $\tau$) are shown in the right corner of the plots. For some stars two models at different neutron densities result in similarly good fits with their respective $\chi^2$ values varying by less than 10\%, in which case we provide both fits and $\chi^2$ values (see also discussions in section \S\ref{sec:comparison}).
The lower panel shows the distribution
of the residuals. The uncertainty of the observations $\sigma_{Z, obs}$ is indicated by errorbars in the upper panel and by the
shaded region in the lower panel. The vertical lines show the location of Sr, Ba and Eu which are representatives of
the ls and hs peak as well as the r process, respectively.
For comparison the best fit from \citet{Abate2015b} is shown if available, which shows the result of binary evolution and AGB nucleosynthesis. We refer the reader to the corresponding figures of \citet{Bisterzo2012} for comparison to s-process models with initial r-process enhancement.

\begin{figure}
\begin{minipage}[b]{.45\textwidth}
\centering
\includegraphics[ width=\linewidth]{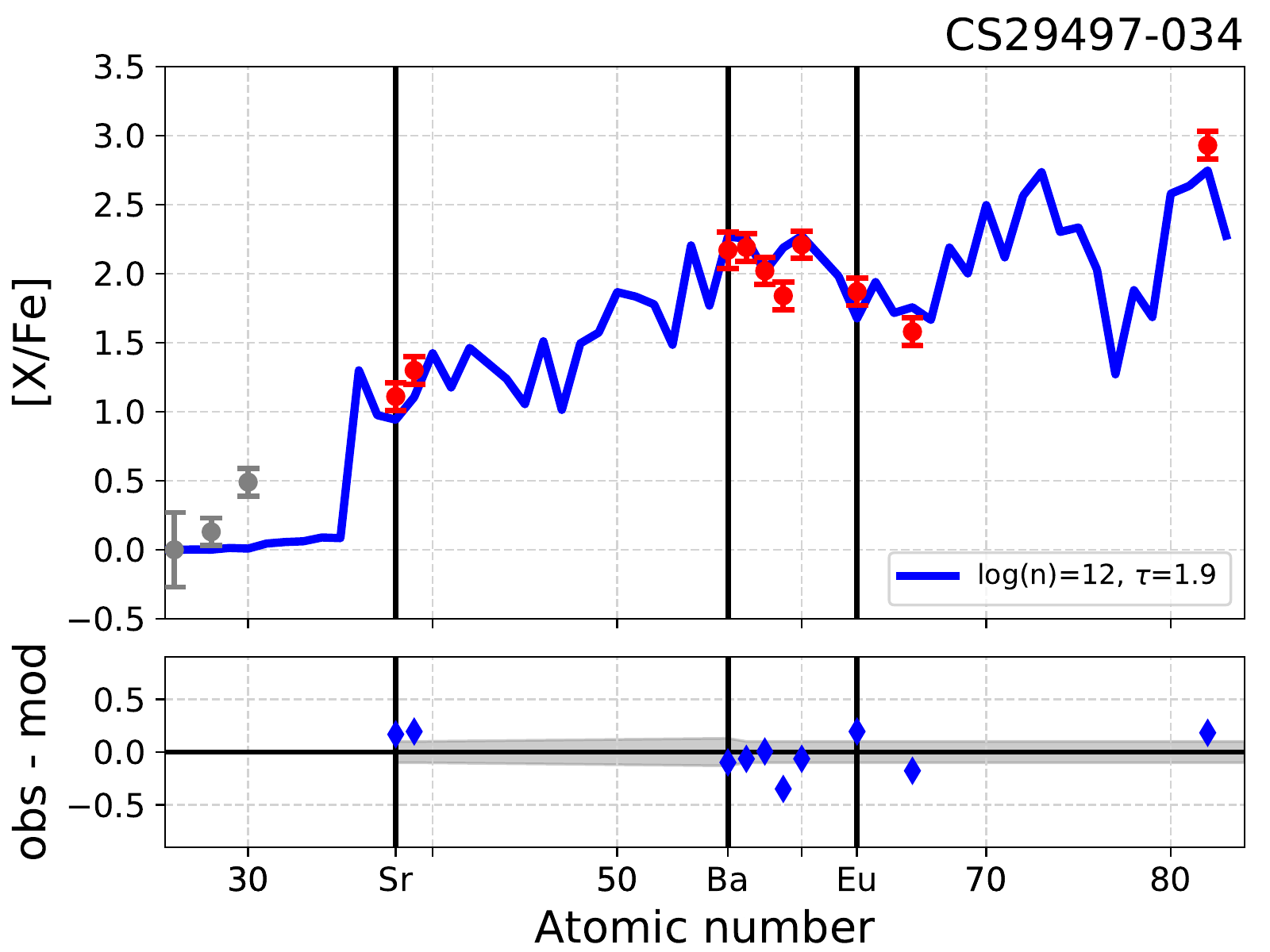}
\caption{Best fitting model for CEMP-i star CS29497-034. The best fitting s-process models with initial r-process enhancement can be found in Fig. 24 of \citet{Bisterzo2012}.}
\end{minipage}
\hfill
\begin{minipage}[b]{.45\textwidth}
\centering
\includegraphics[ width=\linewidth]{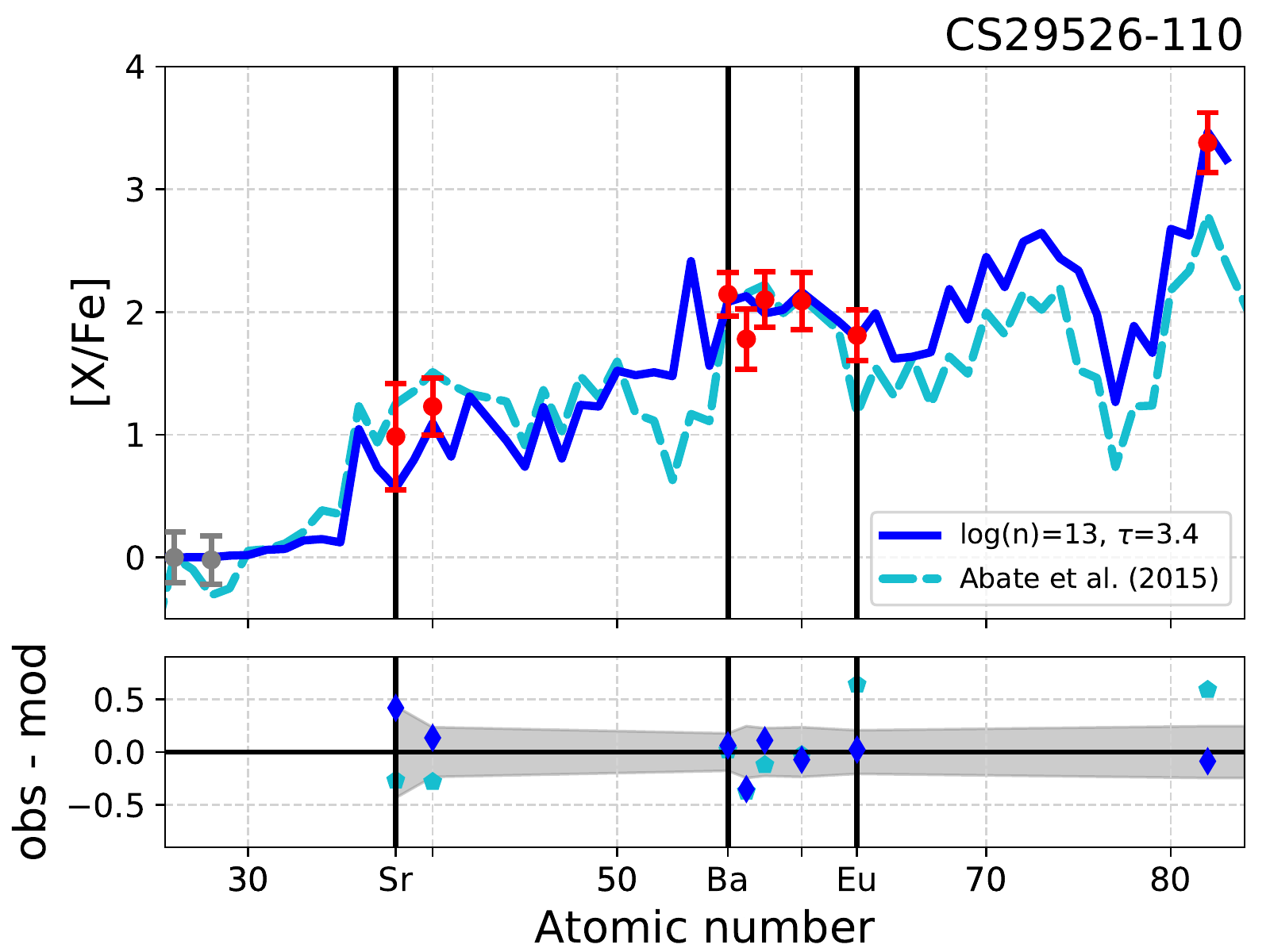}
\caption{Best fitting model for CEMP-i star CS29526-110 compared to the best fit of s-process nucleosynthesis with binary evolution of \citet{Abate2015b}. The best fitting s-process models with initial r-process enhancement can be found in Fig. 23 of \citet{Bisterzo2012}.}
\end{minipage}
\hfill
\begin{minipage}[b]{.45\textwidth}
\centering
\includegraphics[ width=\linewidth]{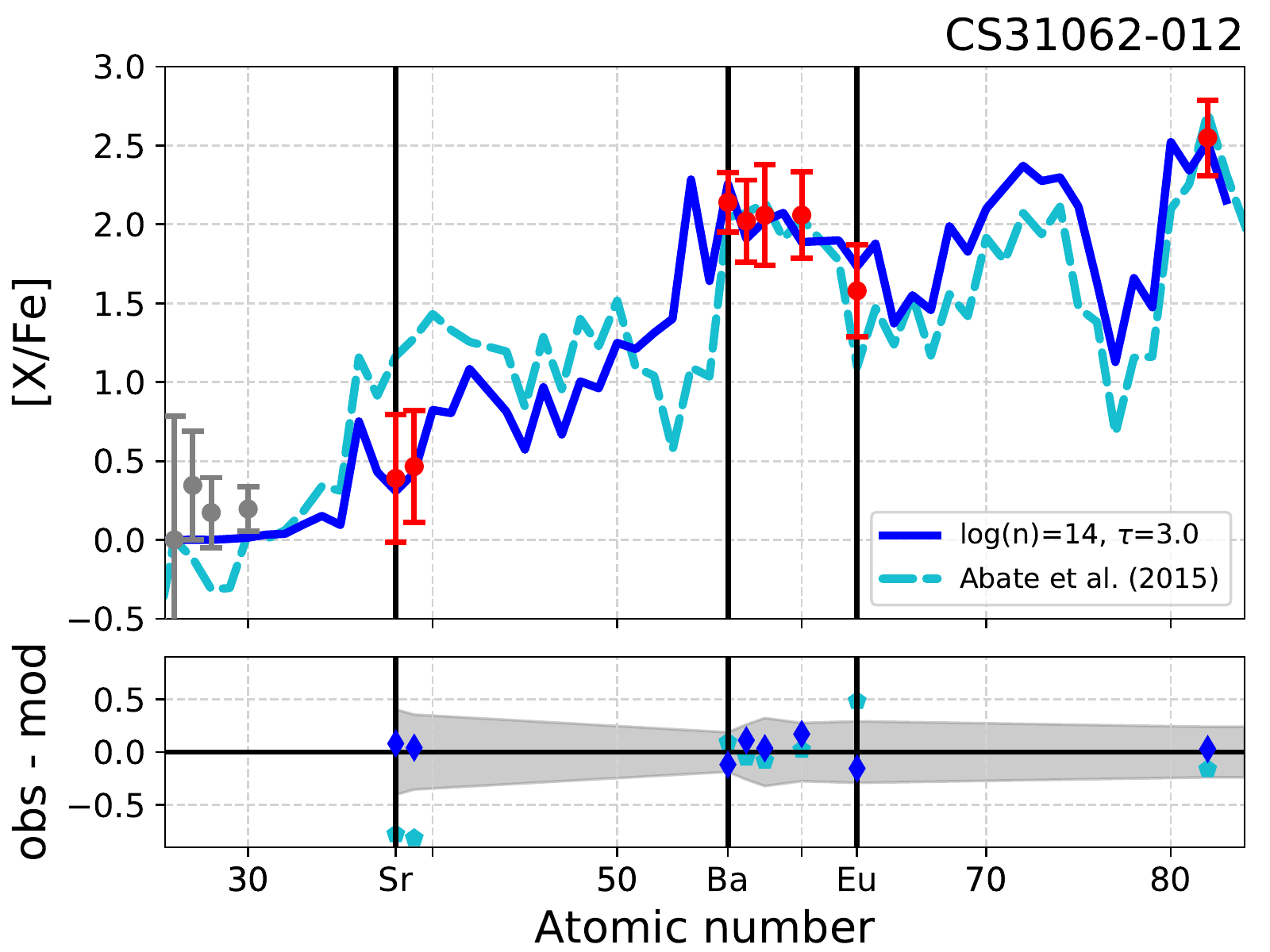}
\caption{Best fitting model for CEMP-i star CS31062-012 compared to the best fit of s-process nucleosynthesis with binary evolution of \citet{Abate2015b}. The best fitting s-process models with initial r-process enhancement can be found in Fig. 24 of \citet{Bisterzo2012}.}
\end{minipage}
\hfill
\begin{minipage}[b]{.45\textwidth}
\centering
\includegraphics[ width=\linewidth]{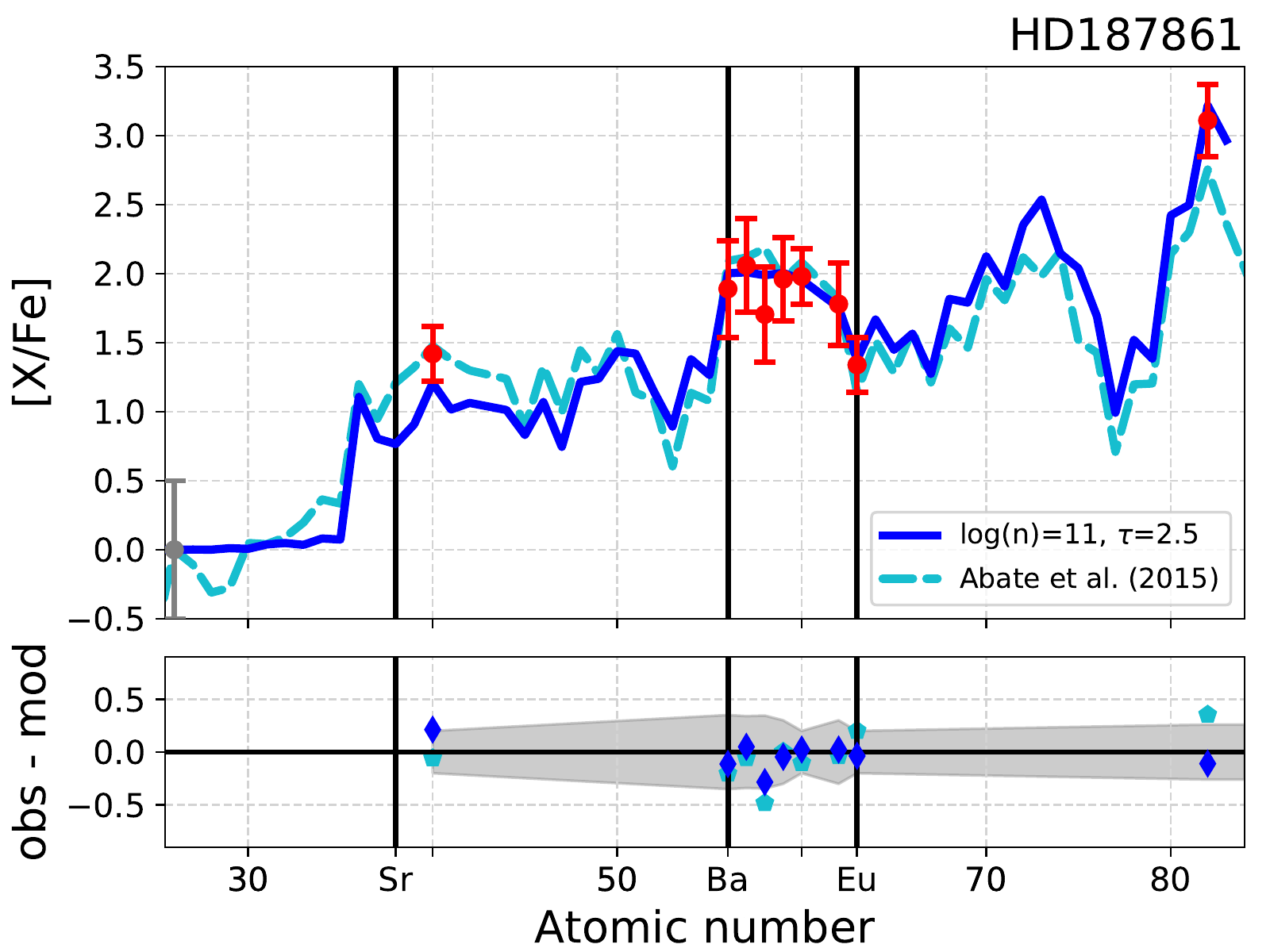}
\caption{Best fitting model for CEMP-i star HD187861 compared to the best fit of s-process nucleosynthesis with binary evolution of \citet{Abate2015b}. The best fitting s-process models with initial r-process enhancement can be found in Fig. 29 of \citet{Bisterzo2012}.}
\end{minipage}
\hfill
\end{figure}
\begin{figure}
\begin{minipage}[b]{.45\textwidth}
\centering
\includegraphics[ width=\linewidth]{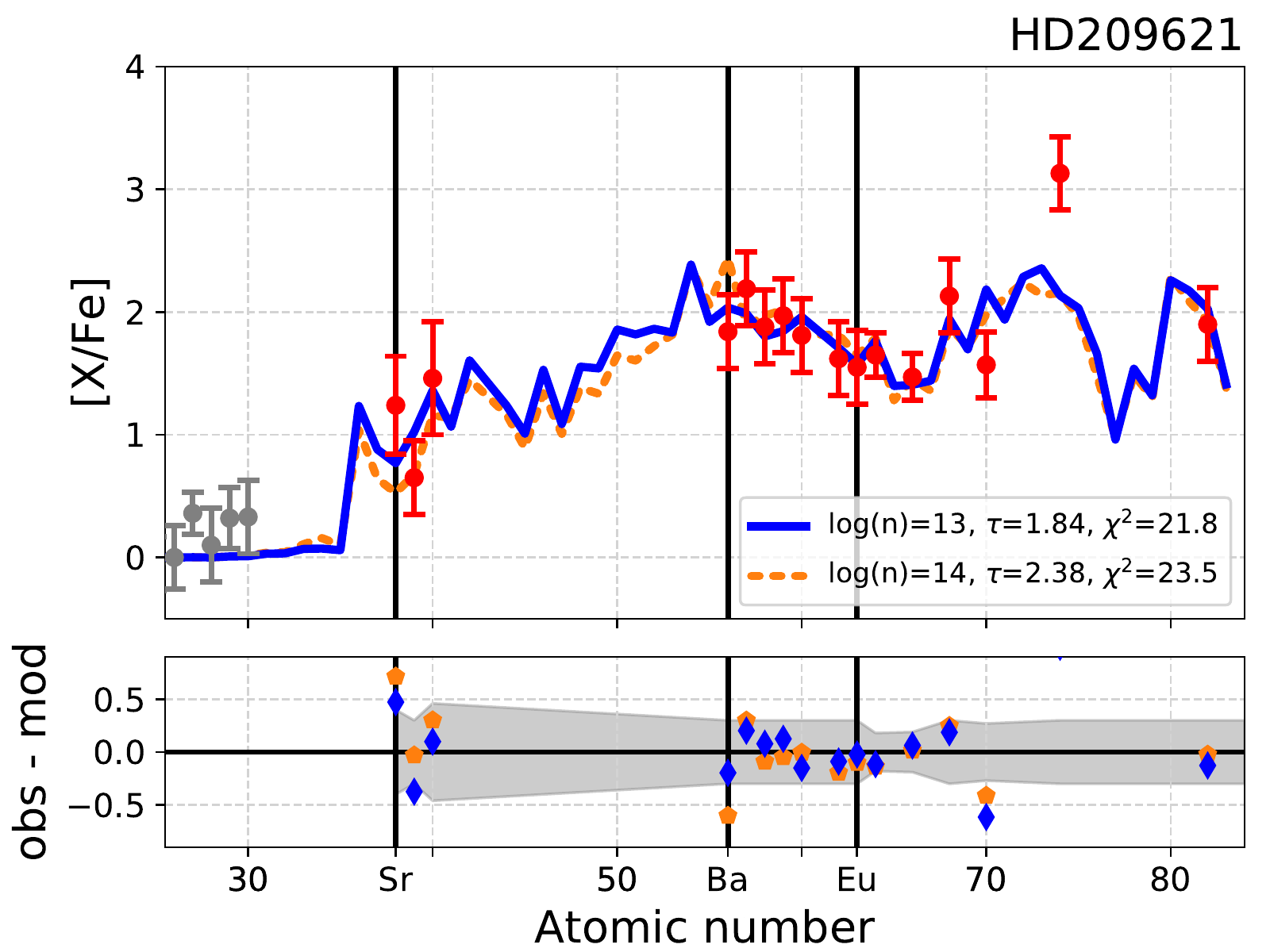}
\caption{Best fitting models for CEMP-i star HD209621. The best fitting s-process models with initial r-process enhancement can be found in Fig. 35 of \citet{Bisterzo2012}.}
\end{minipage}
\hfill
\begin{minipage}[b]{.45\textwidth}
\centering
\includegraphics[ width=\linewidth]{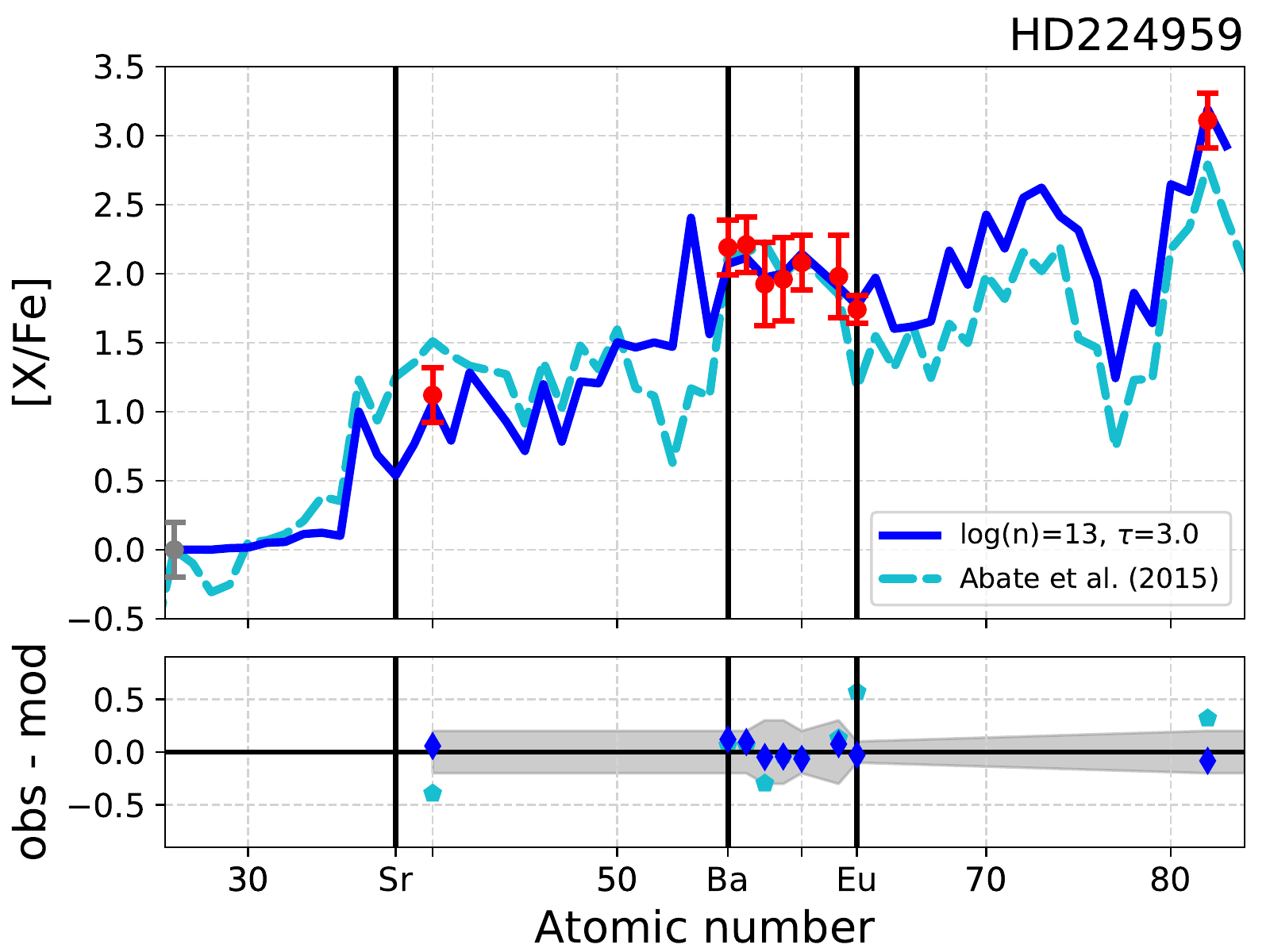}
\caption{Best fitting model for CEMP-i star HD224959 compared to the best fit of s-process nucleosynthesis with binary evolution of \citet{Abate2015b}. The best fitting s-process models with initial r-process enhancement can be found in Fig. 30 of \citet{Bisterzo2012}.}
\end{minipage}
\hfill
\begin{minipage}[b]{.45\textwidth}
\centering
\includegraphics[ width=\linewidth]{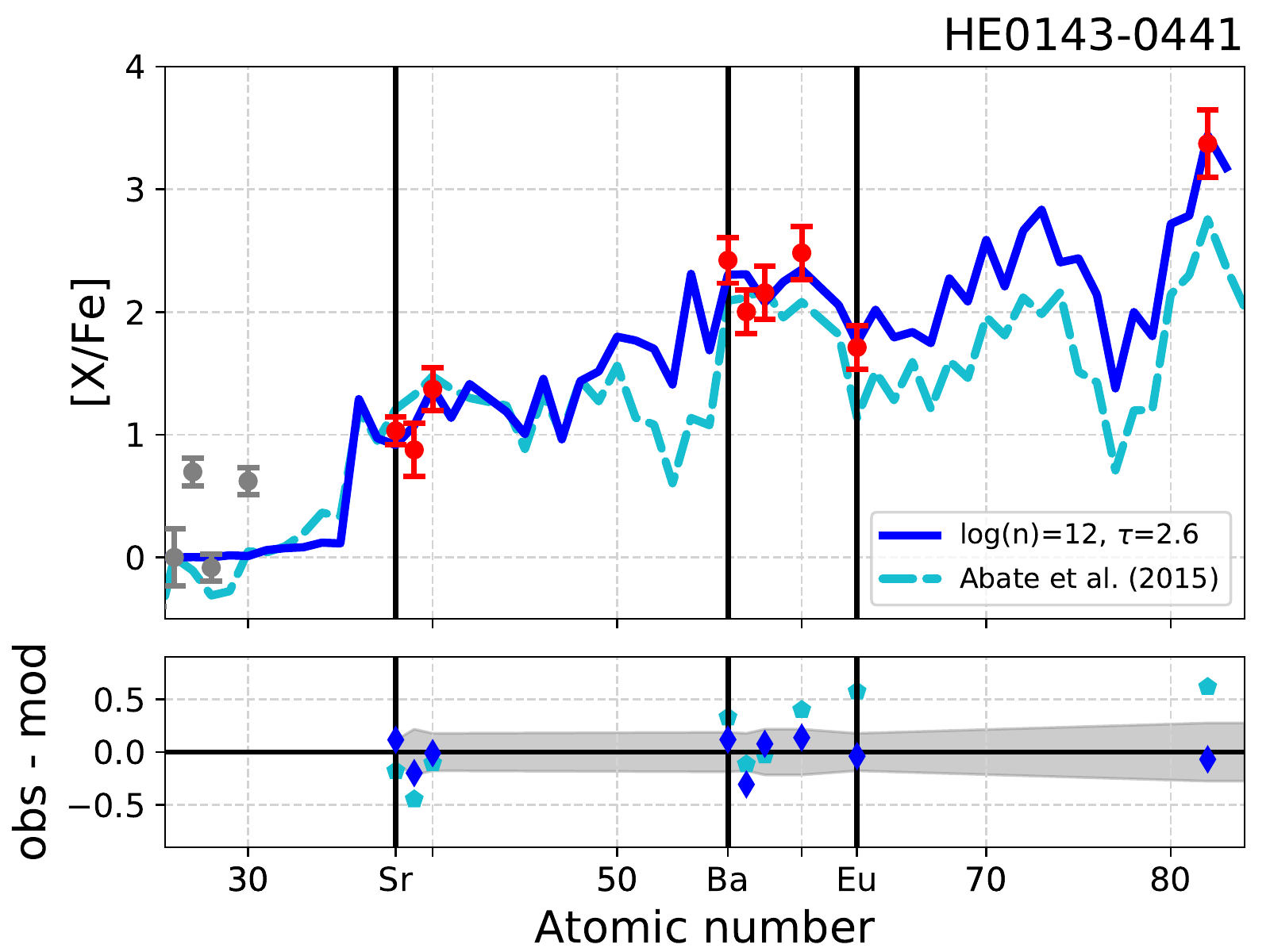}
\caption{Best fitting model for CEMP-i star HE0143-0441 compared to the best fit of s-process nucleosynthesis with binary evolution of \citet{Abate2015b}. The best fitting s-process models with initial r-process enhancement can be found in Fig. 33 of \citet{Bisterzo2012}.}
\end{minipage}
\hfill
\begin{minipage}[b]{.45\textwidth}
\centering
\includegraphics[ width=\linewidth]{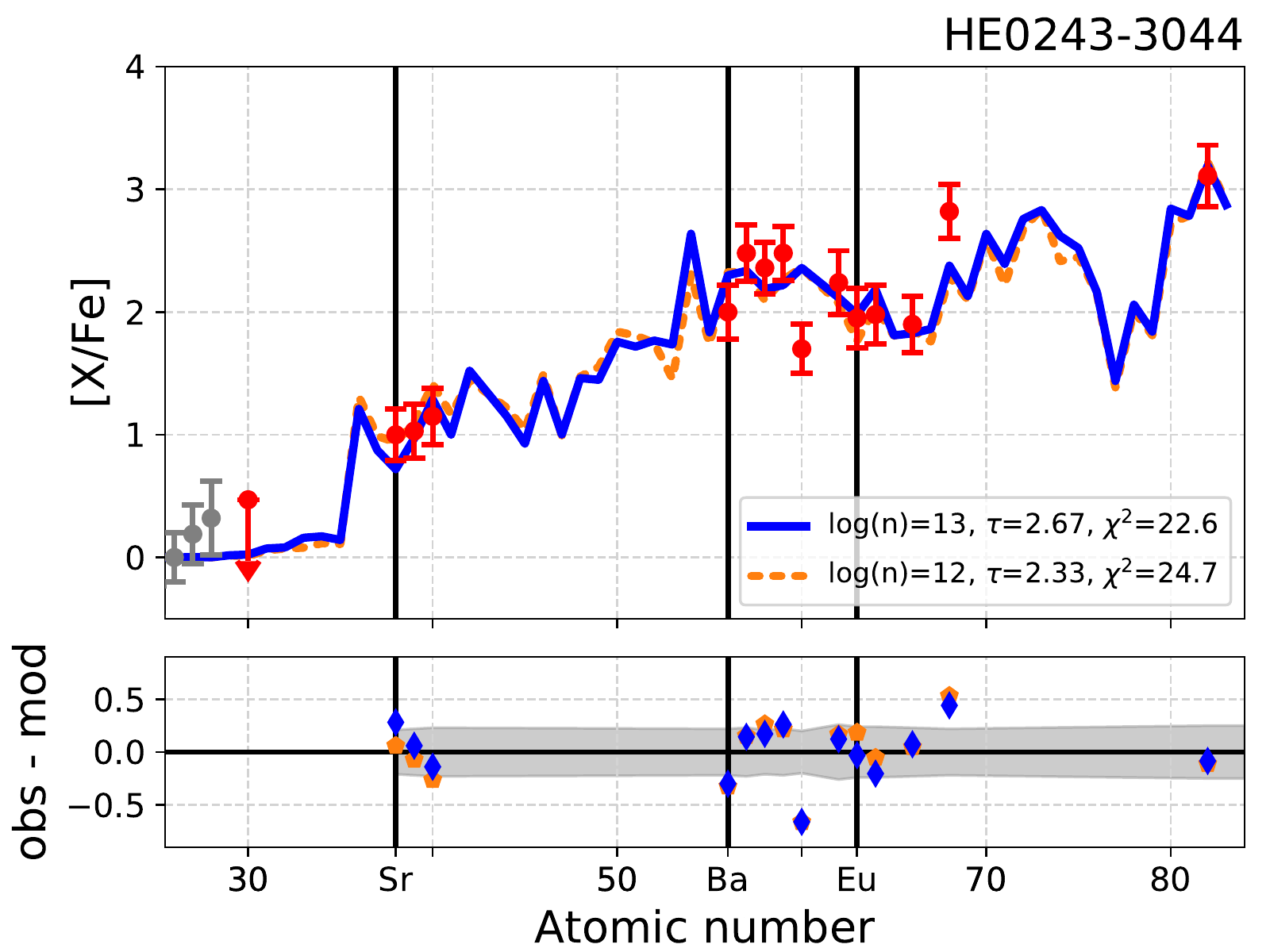}
\caption{Best fitting models for CEMP-i star HE0243-3044.}
\end{minipage}
\hfill
\begin{minipage}[b]{.45\textwidth}
\centering
\includegraphics[ width=\linewidth]{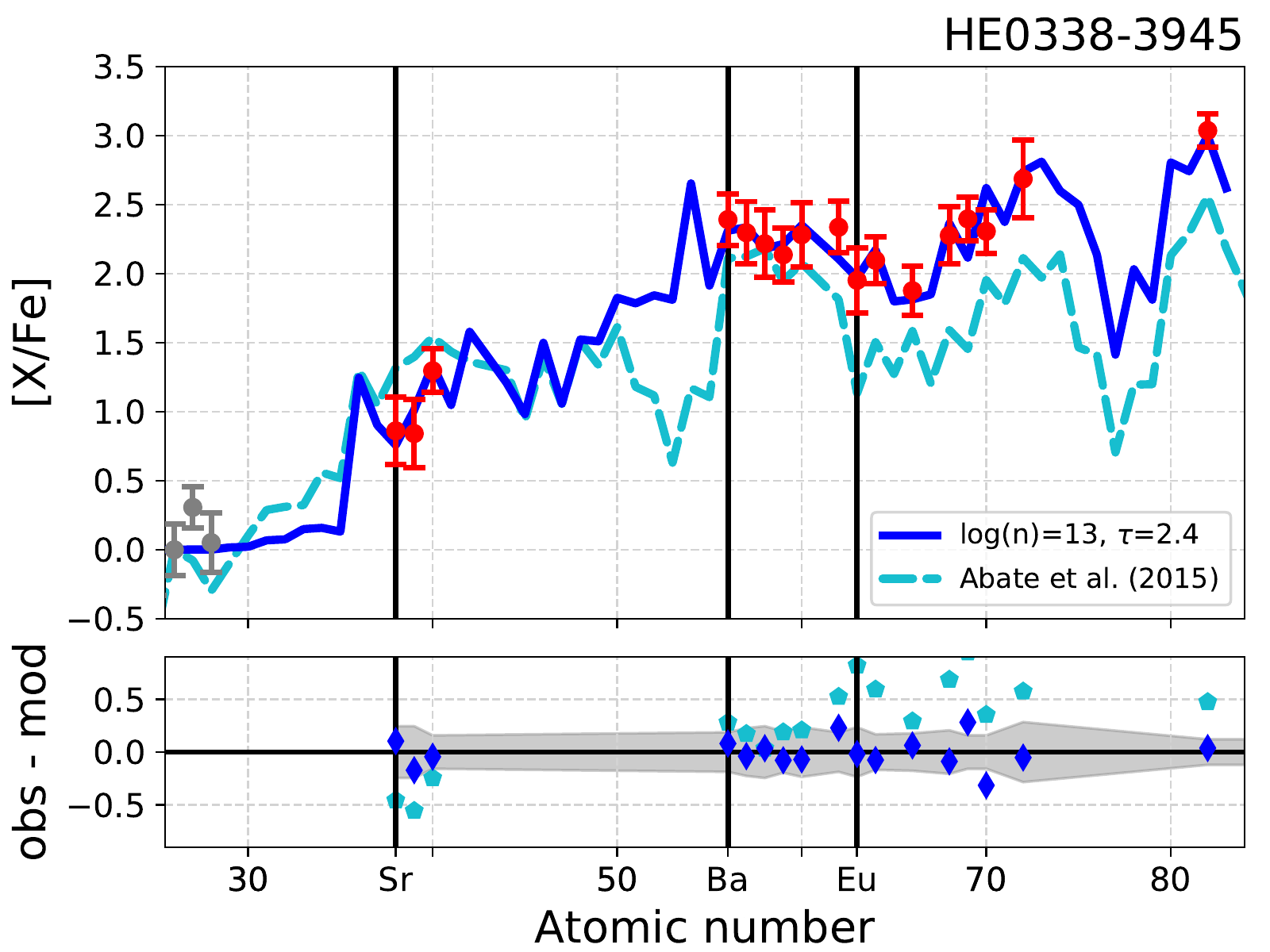}
\caption{Best fitting model for CEMP-i star HE0338-3945 compared to the best fit of s-process nucleosynthesis with binary evolution of \citet{Abate2015b}. The best fitting s-process models with initial r-process enhancement can be found in Fig. 19 of \citet{Bisterzo2012}.}
\end{minipage}
\hfill
\end{figure}
\begin{figure}
\begin{minipage}[b]{.45\textwidth}
\centering
\includegraphics[ width=\linewidth]{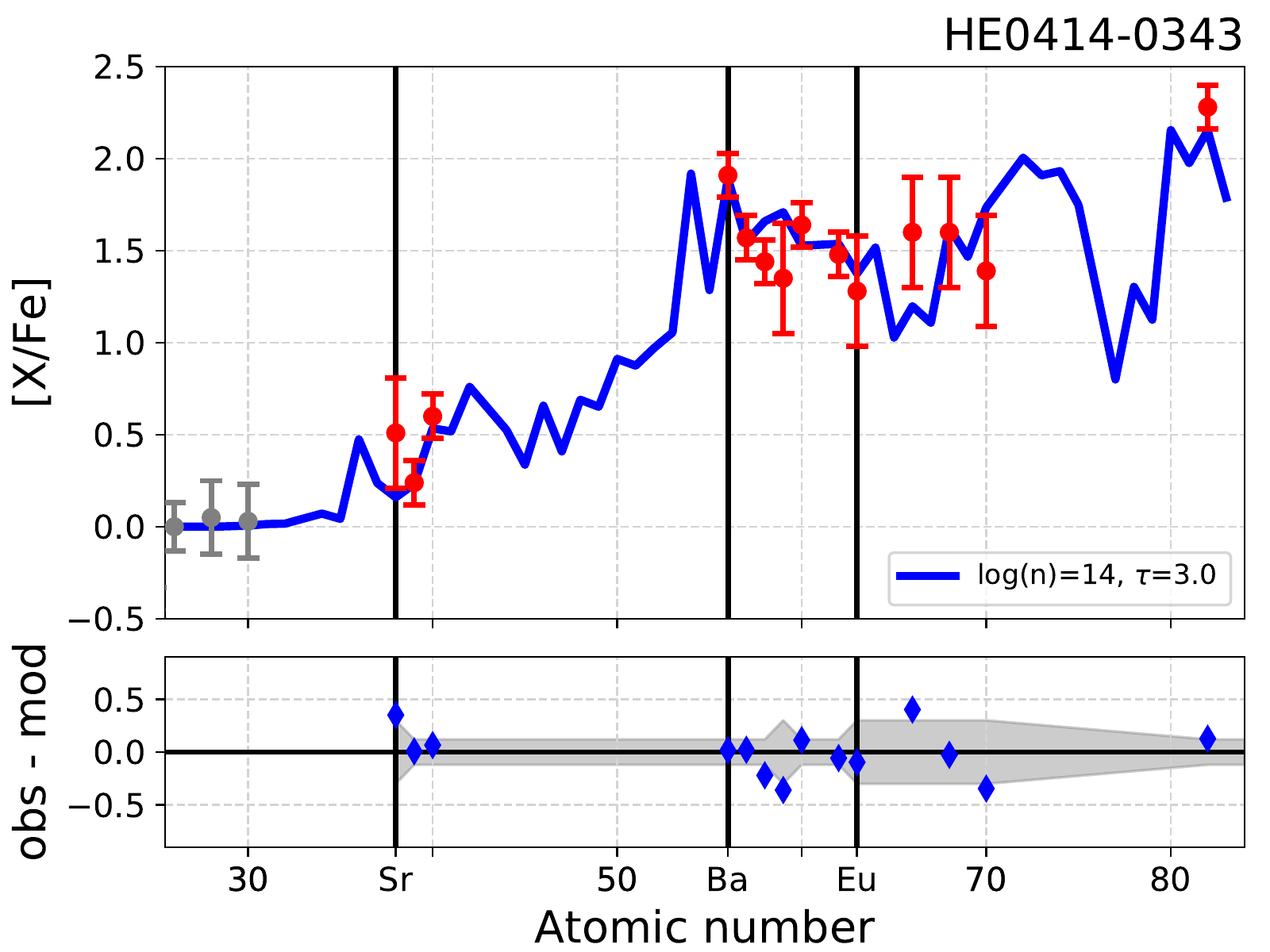}
\caption{Best fitting model for CEMP-i star HE0414-0343.}
\end{minipage}
\hfill
\begin{minipage}[b]{.45\textwidth}
\centering
\includegraphics[ width=\linewidth]{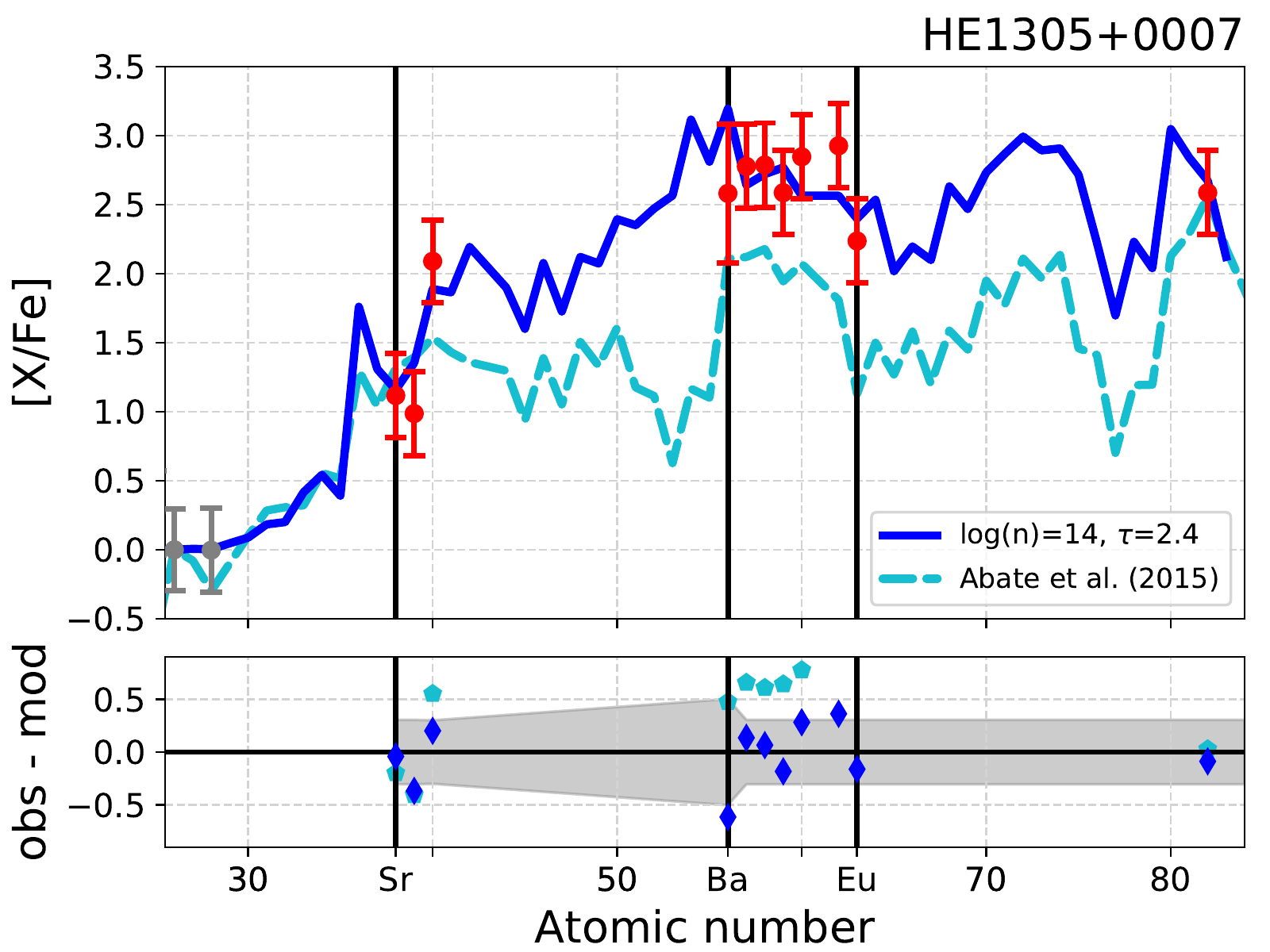}
\caption{Best fitting model for CEMP-i star HE1305+0007 compared to the best fit of s-process nucleosynthesis with binary evolution of \citet{Abate2015b}. The best fitting s-process models with initial r-process enhancement can be found in Fig. 22 of \citet{Bisterzo2012}.}
\end{minipage}
\hfill
\begin{minipage}[b]{.45\textwidth}
\centering
\includegraphics[ width=\linewidth]{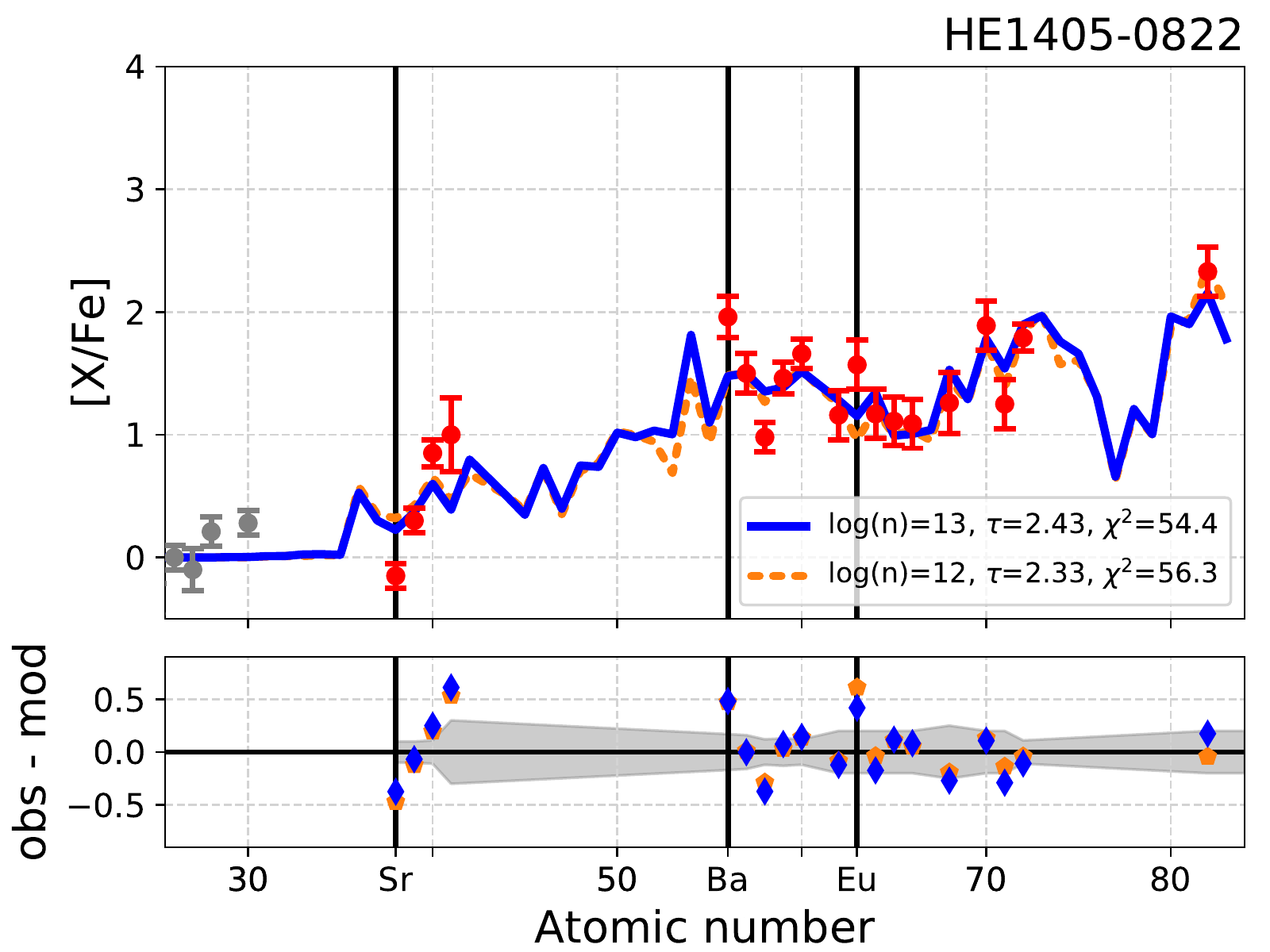}
\caption{Best fitting models for CEMP-i star HE1405-0822.}
\end{minipage}
\hfill
\begin{minipage}[b]{.45\textwidth}
\centering
\includegraphics[ width=\linewidth]{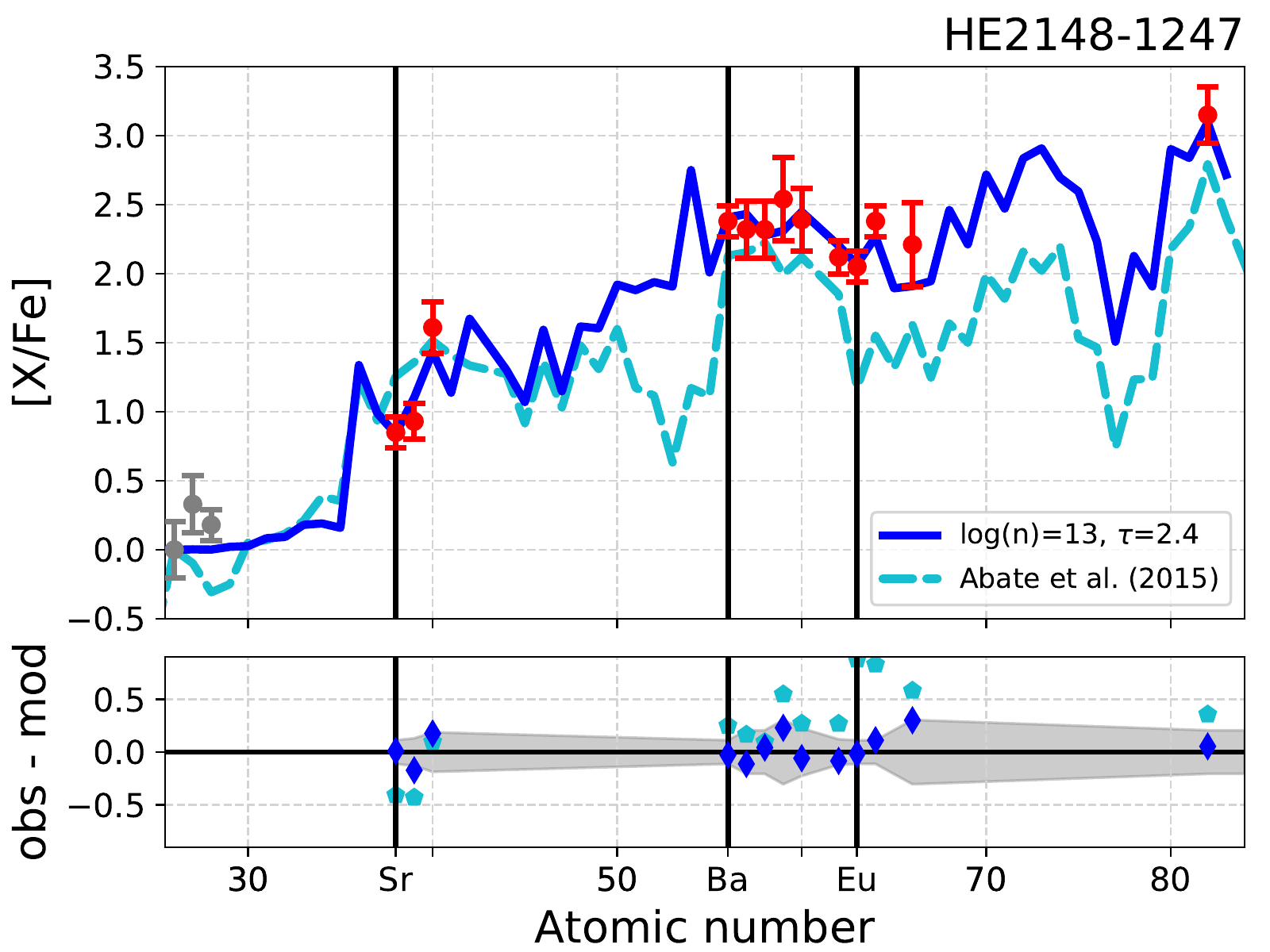}
\caption{Best fitting model for CEMP-i star HE2148-1247 compared to the best fit of s-process nucleosynthesis with binary evolution of \citet{Abate2015b}. The best fitting s-process models with initial r-process enhancement can be found in Fig. 21 of \citet{Bisterzo2012}.}\label{Fig:app_cemp_last}
\end{minipage}
\end{figure}

\section{All fits: post-AGB stars (online only)} \label{app:pAGB_fits}
This section shows the best fitting models for each of the 7 post-AGB stars in comparison to the observed abundance patterns in Figures \ref{Fig:app_pagb_first} to \ref{Fig:app_pagb_last}. As in Appendix \ref{app:cemp_fits}, 
details of each best fit (neutron
density $n$ and time-integrated neutron exposure $\tau$) are shown in the right corner of the plots. The lower panel shows the distribution
of the residuals. The uncertainty of the observations $\sigma_{Z, obs}$ is indicated by errorbars in the upper panel and by the
shaded region in the lower panel. The vertical lines show the location of Sr, Ba and Eu which are representatives of
the ls and hs peak as well as the r process, respectively.
For comparison the fits of pocket case 3 and 4 from \citet{Lugaro2015} are shown if available. These fits show the result of AGB nucleosynthesis with modified \el{13}{C} pockets to not overproduce the upper limit of the Pb abundance.

\begin{figure}[h]
\begin{minipage}[b]{.45\textwidth}
\centering
\includegraphics[ width=\linewidth]{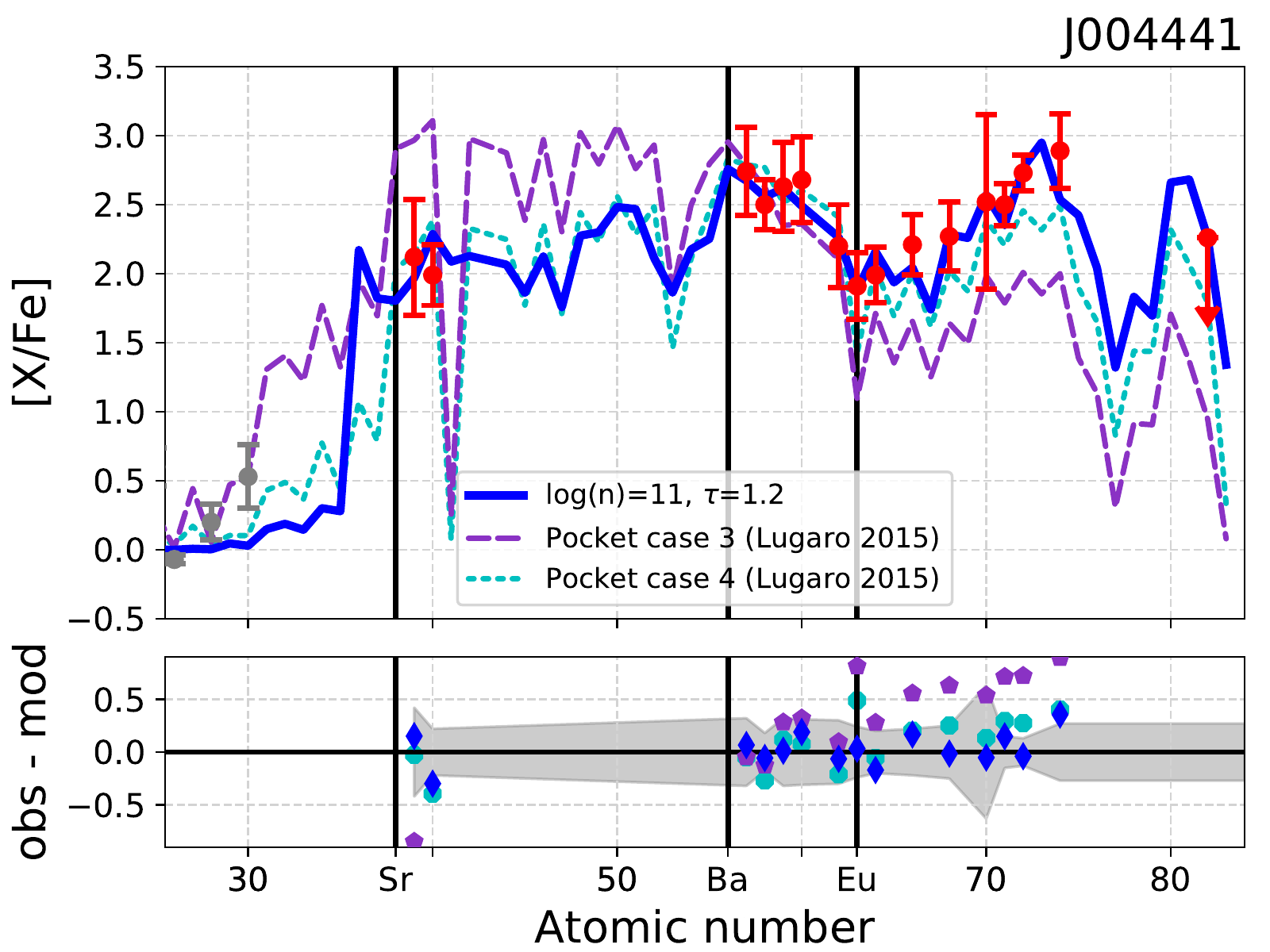}
\caption{Best fitting model for post-AGB star J004441. The dashed and dotted lines show the fits of pocket case 3 and 4 from \citet{Lugaro2015}, which represent AGB nucleosynthesis with modified \el{13}{C} pockets.}\label{Fig:app_pagb_first}
\end{minipage}
\hfill
\begin{minipage}[b]{.45\textwidth}
\centering
\includegraphics[ width=\linewidth]{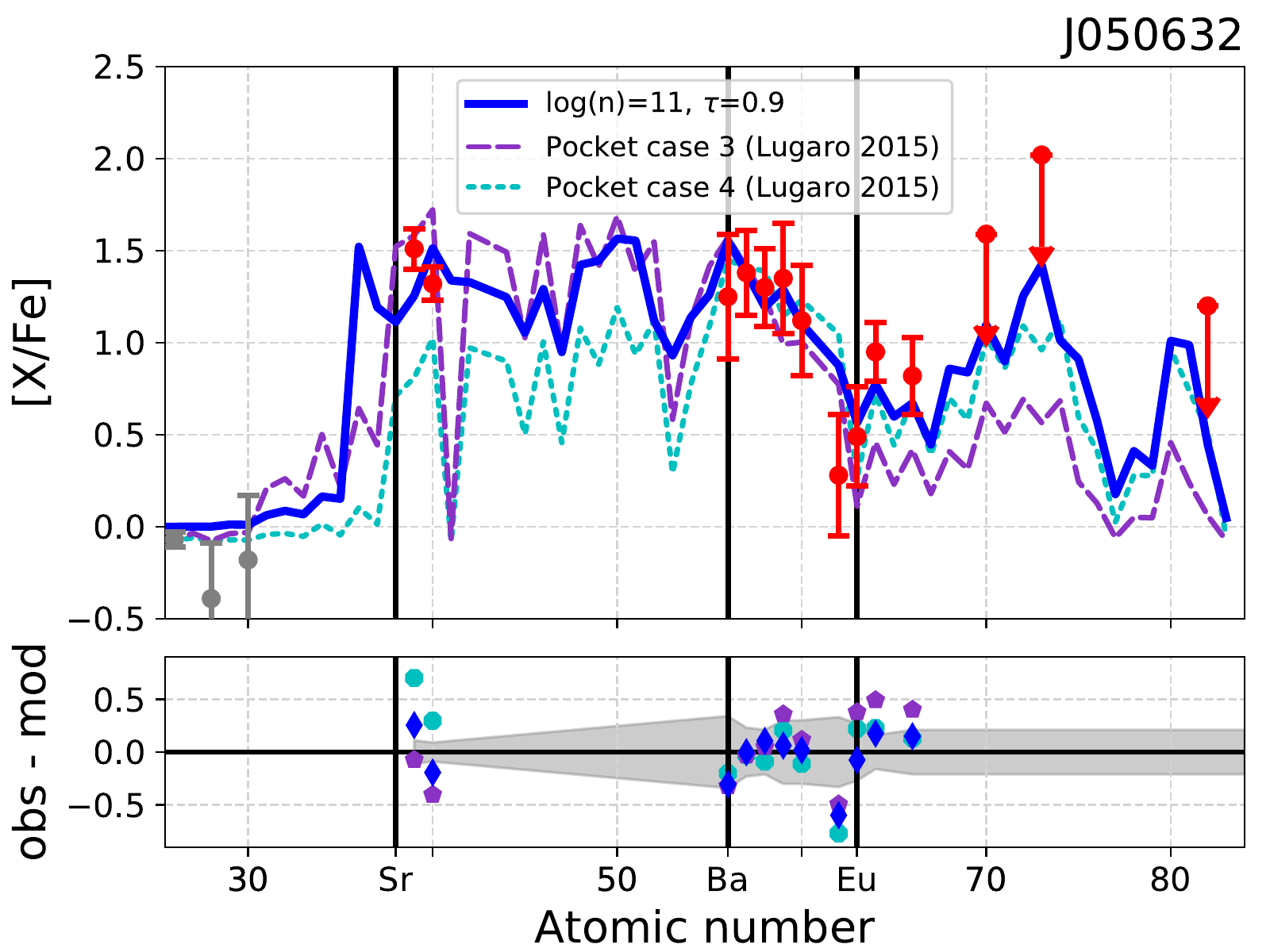}
\caption{Best fitting model for post-AGB star J050632. The dashed and dotted lines show the fits of pocket case 3 and 4 from \citet{Lugaro2015}, which represent AGB nucleosynthesis with modified \el{13}{C} pockets.}
\end{minipage}
\hfill
\begin{minipage}[b]{.45\textwidth}
\centering
\includegraphics[ width=\linewidth]{J052043.pdf}
\caption{Best fitting model for post-AGB star J052043. The dashed and dotted lines show the fits of pocket case 3 and 4 from \citet{Lugaro2015}, which represent AGB nucleosynthesis with modified \el{13}{C} pockets.}
\end{minipage}
\hfill
\begin{minipage}[b]{.45\textwidth}
\centering
\includegraphics[ width=\linewidth]{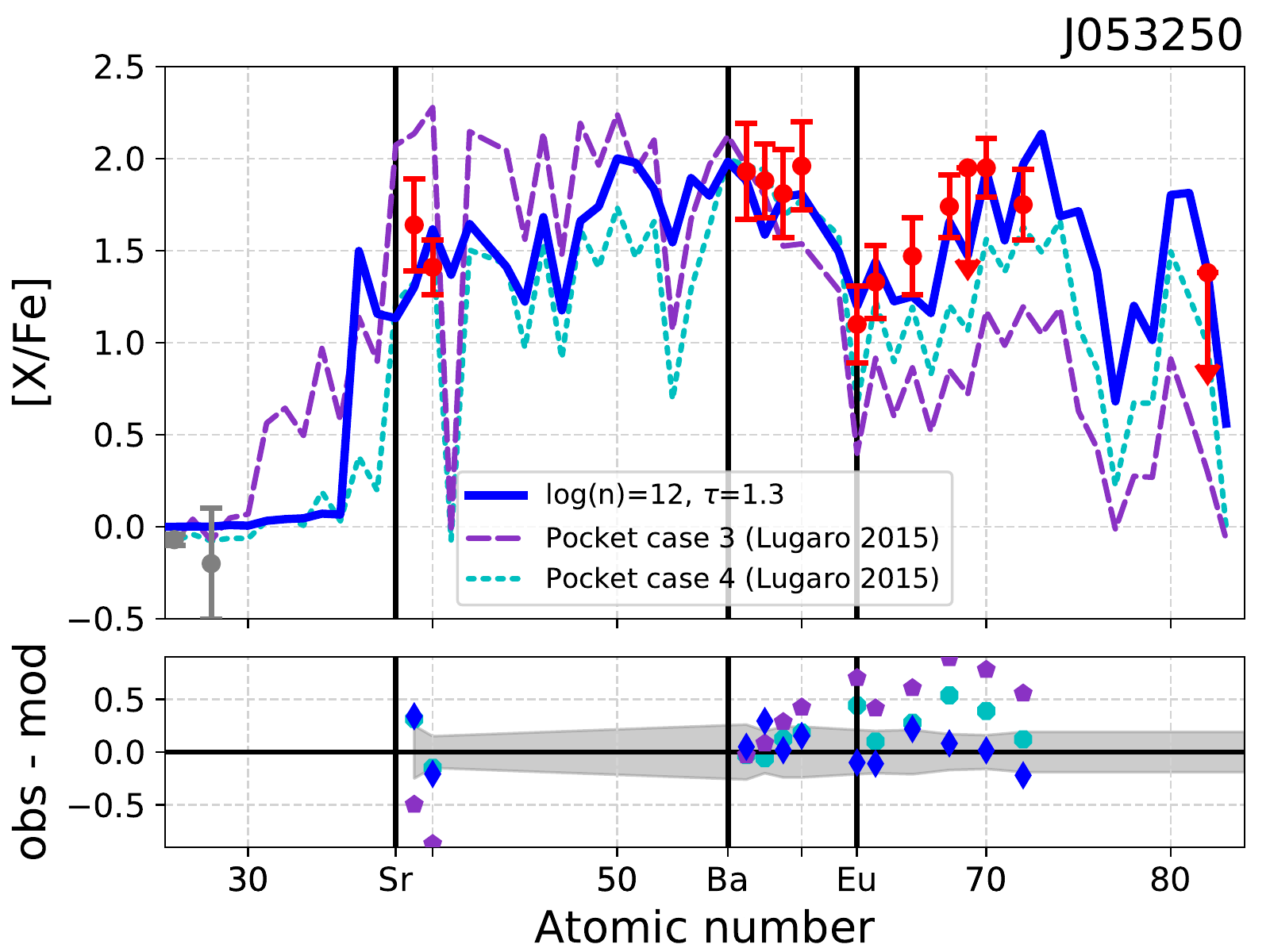}
\caption{Best fitting model for post-AGB star J053250. The dashed and dotted lines show the fits of pocket case 3 and 4 from \citet{Lugaro2015}, which represent AGB nucleosynthesis with modified \el{13}{C} pockets.}
\end{minipage}
\hfill
\begin{minipage}[b]{.45\textwidth}
\centering
\includegraphics[ width=\linewidth]{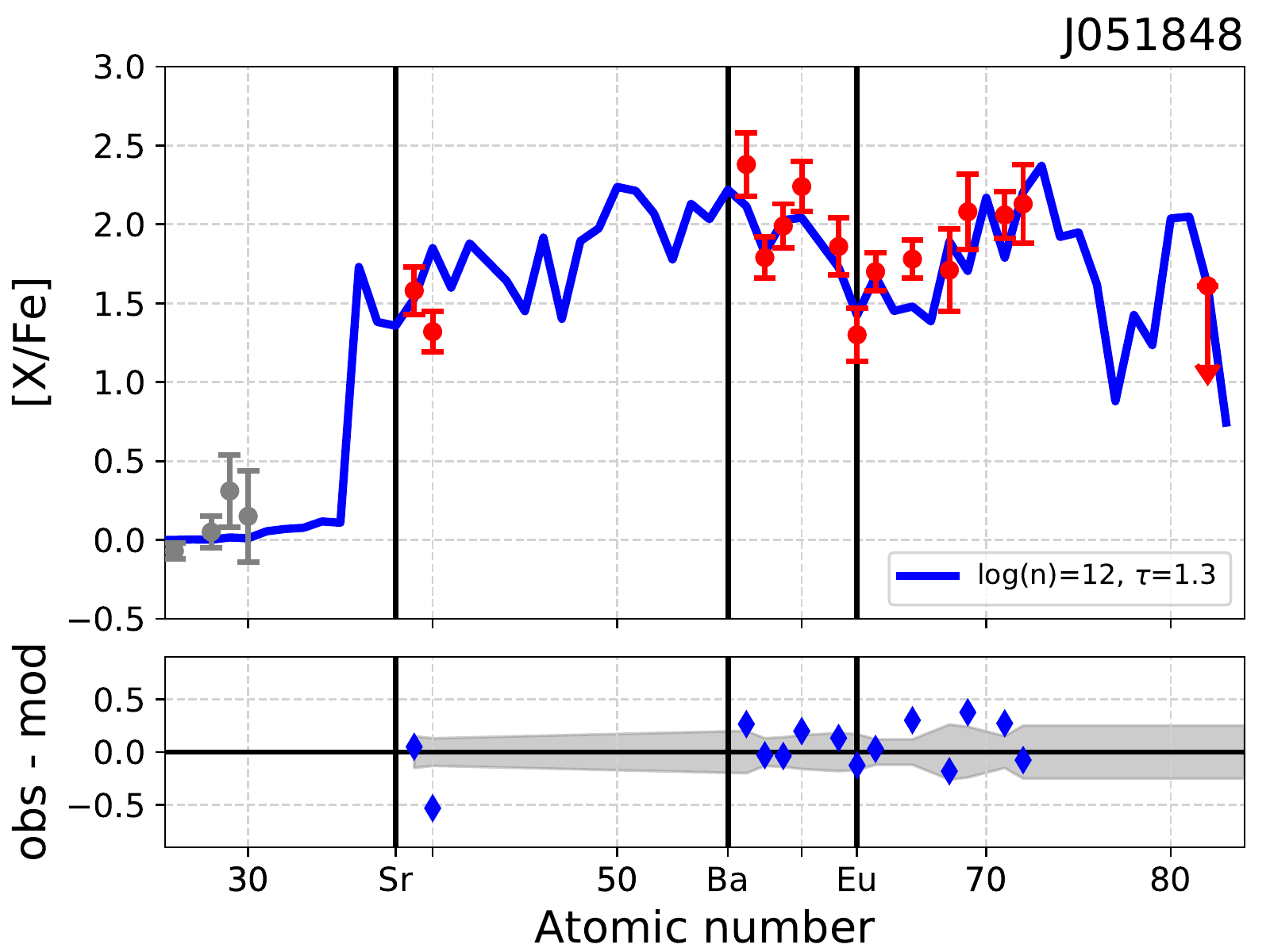}
\caption{Best fitting model for post-AGB star J051848.}
\end{minipage}
\hfill
\begin{minipage}[b]{.45\textwidth}
\centering
\includegraphics[ width=\linewidth]{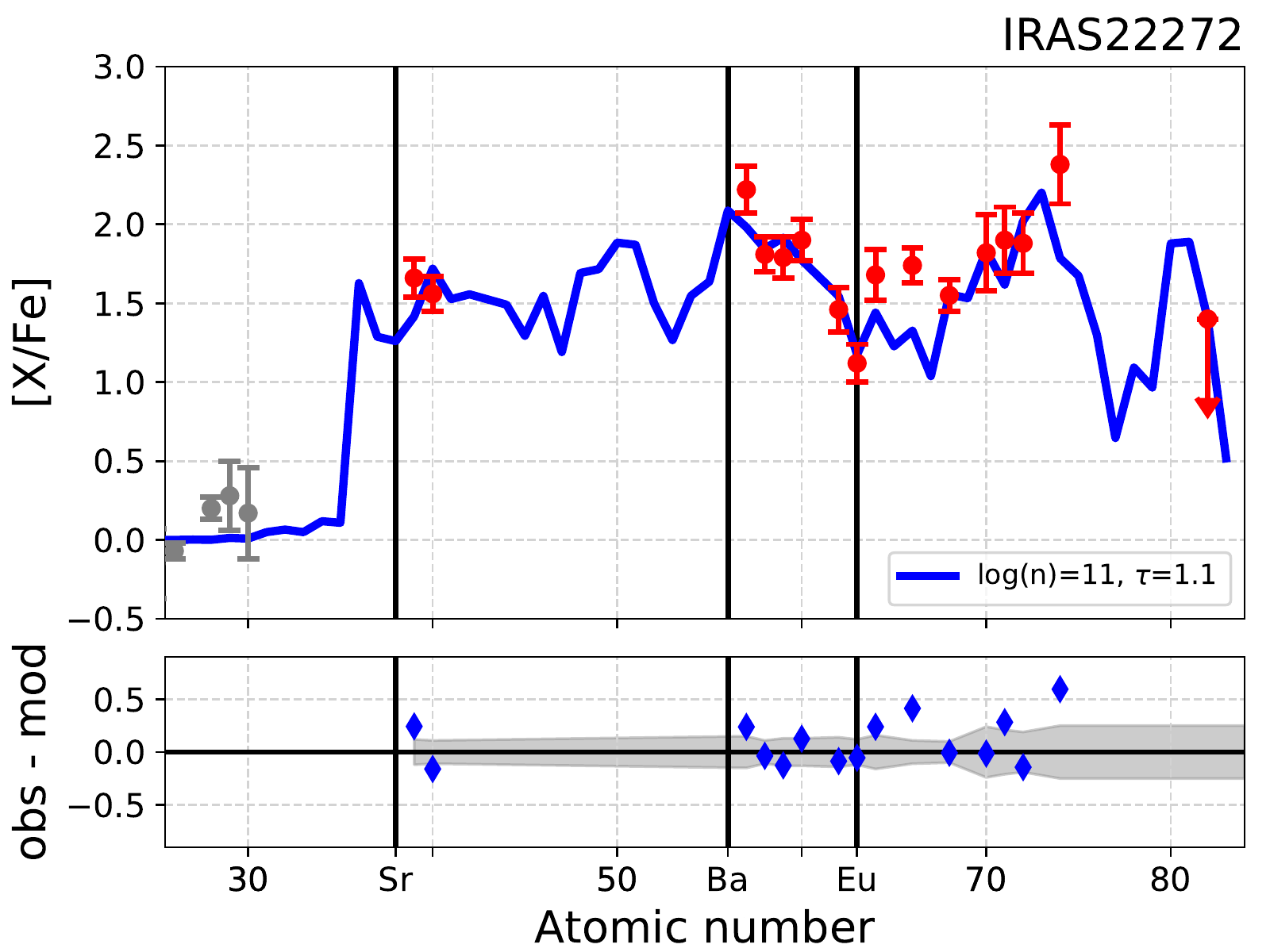}
\caption{Best fitting model for post-AGB star IRAS22272.}
\end{minipage}
\end{figure}
\begin{figure}

\quad

\begin{minipage}[b]{.45\textwidth}
\centering
\includegraphics[ width=\linewidth]{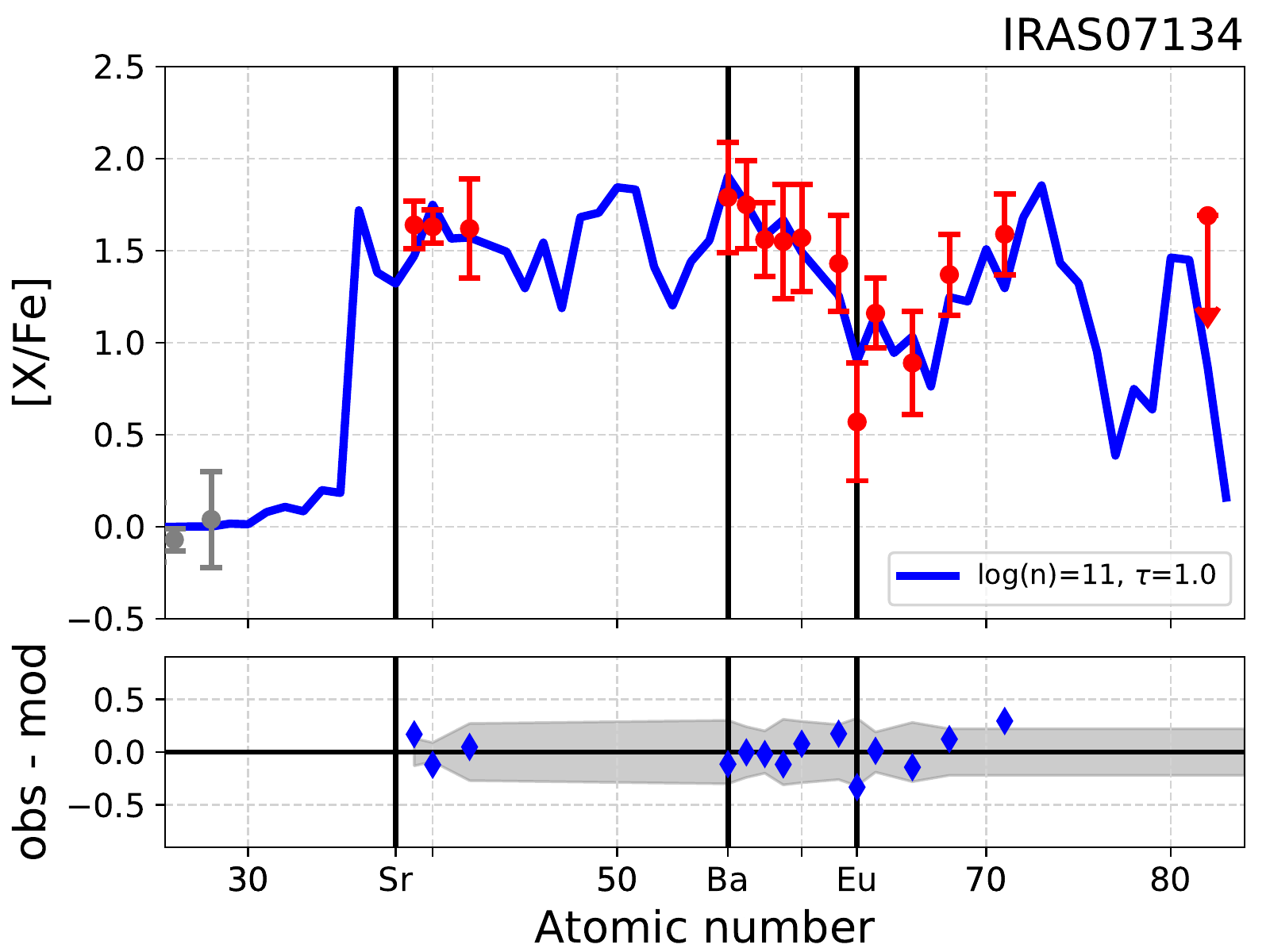}
\caption{Best fitting model for post-AGB star IRAS07134.} \label{Fig:app_pagb_last}
\end{minipage}
\end{figure}

\end{document}